\documentclass[10pt]{iopart}

\usepackage{iopams}

\begin{document}

\title[The Newtonian limit of metric gravity theories with
quadratic Lagrangians]{The Newtonian limit of metric gravity
theories with quadratic Lagrangians}

\author{S. Capozziello$^1$, A. Stabile$^2$}
\address{$^1$Dipartimento di Scienze Fisiche and INFN, Sez. di Napoli, Universit\'a di Napoli ``Federico II'',
Compl. Univ. di Monte S. Angelo, Ed. G., Via Cinthia, I-80126
Napoli, Italy}\ead{capozziello@na.infn.it}
\address{$^3$ Dipartimento di Ingegneria, Universit\'a del Sannio,
 Benevento, C.so
Garibaldi 107, I-80125 Benevento, Italy}\ead{arturo.stabile@sa.infn.it}

\begin{abstract}
The Newtonian limit of  fourth-order gravity  is worked out
discussing its viability with respect to the standard results of
General Relativity. We  investigate the limit in the metric
approach which, with respect to the  Palatini formulation, has
been much less studied in the recent literature, due to the
higher-order of the field equations. In addition, we refrain from
exploiting the formal equivalence of higher-order theories
considering the analogy with specific  scalar-tensor theories,
i.e. we work in the so-called Jordan frame in order to avoid
possible misleading interpretations of the results. Explicit
solutions are provided for several different types of Lagrangians
containing powers of the Ricci scalar as well as combinations of
the other curvature invariants. In particular, we develop the
Green function method for fourth-order theories in order to find
out solutions. Finally, the consistency of the results with
respect to General Relativity is discussed.
\end{abstract}

\pacs{04.25.-g; 04.25.Nx; 04.40.Nr} \vspace{2pc} \noindent{\it
Keywords}: Alternative gravity theories; perturbation theory;
Newtonian approximation

\maketitle

%% 04.25.-g Approximation methods; equations of motion
%% 04.25.Nx Post-Newtonian approximation; perturbation theory; related approximations
%% 04.40.Nr Einstein-Maxwell spacetimes, spacetimes with fluids, radiation or classical fields
%% 98.80.Jk Mathematical and relativistic aspects of cosmology

\maketitle

\section{Introduction}

The study of possible modifications of Einstein's theory of
gravity has a long history which reaches back to the early 1920s
\cite{Weyl:1918,Pauli:1919,Bach:1921}.

Corrections to the gravitational Lagrangian, leading to
higher-order field equations, were already studied by several
authors \cite{Weyl:1921,Eddington:1924,Lanczos:1931} shortly after
General Relativity was proposed. Developments in the 1960s and
1970s
\cite{Buchdahl:1962,DeWitt:1965,Bicknell:1974,Havas:1977,Stelle:1978},
partly motivated by the quantization schemes proposed at that
time, made clear that theories containing {\it only} a $R^2$ term
in the Lagrangian were not viable with respect to their weak field
behavior. Buchdahl, in 1962 \cite{Buchdahl:1962}, rejected pure
$R^2$ theories because of the non-existence of asymptotically flat
solutions.

The early proposed amendments of Einstein's theory were aimed at a
unification of gravity with other branches of physics, like
Electromagnetism; recently the interest in such modifications
comes also from cosmology. For a comprehensive review, see
\cite{Schmidt:2004}. In order to explain observational data,
additional ad-hoc concepts, like dark energy/matter, are
introduced within Einstein's theory. On the other hand, the
emergence of such stopgap measures in a cosmological context could
be interpreted as a first signal of the breakdown of General
Relativity on these scales \cite{JCAP,MNRAS}, and led to the
proposal of many alternative modifications of the underlying
gravity theory (see \cite{OdiRev,grg} for  reviews).

While it is very natural to extend Einstein's gravity to theories
with additional geometric degrees of freedom, see for example
\cite{Hehl:1976,Hehl:1995,Trautman:2006} for some general surveys
on this subject as well as \cite{Puetzfeld:2005} for a list of
works in a cosmological context, recent attempts focused on the
old idea of modifying the gravitational Lagrangian in a purely
metric framework, leading to higher-order field equations. Due to
the increased complexity of the field equations in this framework,
the main body of works dealt with some formally equivalent
theories, in which a reduction of the order of the field equations
was achieved by considering the metric and the connection as
independent objects \cite{Francaviglia}. In addition, many authors
exploited the formal relationship to scalar-tensor theories to
make some statements about the weak field regime, which was
already worked out for scalar-tensor theories
\cite{Damour:Esposito-Farese:1992}.

In this paper,  we shall study the Newtonian limit of fourth-order
gravity theories in which extensions of the Hilbert-Einstein
Lagrangian are considered.  We are going to focus on the weak
field limit  within the metric approach.  At this point it is
useful to remind  that it was already shown in
\cite{Buchdahl:1979} that different variational procedures do not
lead to equivalent results in the case of quadratic order
Lagrangians.

In addition, we carry out our analysis in the so called the Jordan
frame, i.e.\ we do {\it not} reduce the theory under consideration
to a simpler one by means of a conformal transformation. This is
due to the fact that it was already shown earlier
\cite{Dicke:1962,Brans:1988} that non-linear theories of the kind
considered here, with the exception of some special cases, could
be {\it not} physically equivalent if they have undergone a
conformal transformation. The debate on this topic is open as can
be seen in recent literature (see for example
\cite{Faraoni,CNOT,tsu2,tsu3,noi-phase,ctd}).

By considering some admissible choices for the gravitational
Lagrangian with quadratic corrections, we explicitly work out the
weak field limit. Such considerations are developed also in
relation to the companion papers
\cite{Newton:2007,Curv:2007,noether} where we have considered the
Newtonian limit explicitly for $f(R)$ gravity, taking into account
generic analytic functions of the Ricci scalar $R$, and the
spherically symmetric solutions vs. the weak field limit
respectively.

In principle, any alternative or extended theory of gravity should
allow to recover positive results of General Relativity, for
example in a weak limit regime,  then starting from the
Hilbert-Einstein Lagrangian
\begin{eqnarray}
\mathcal{L}_0 &=& R, \label{eq:generals:Ls:0}
\end{eqnarray}
the following terms
\begin{eqnarray}
\mathcal{L}_1 &=& R^2, \label{eq:generals:Ls:1} \\
\mathcal{L}_2 &=& R_{\alpha \beta} R^{\alpha \beta}, \label{eq:generals:Ls:2} \\
\mathcal{L}_3 &=& R_{\alpha \beta \mu \nu} R^{\alpha \beta \mu
\nu}, \label{eq:generals:Ls:3}
\end{eqnarray}
and combinations of them, represent the obvious minimal choices
for an extended gravity theory  with respect to General
Relativity. Since the variational derivative of $\mathcal{L}_3$
can be linearly expressed
\cite{Bach:1921,Lanczos:1938,Teyssandier:Tourrenc:1983,Kerner} via
the variational derivatives of $\mathcal{L}_1$ and
$\mathcal{L}_2$, one may omit $\mathcal{L}_3$ in the final
Lagrangian of a fourth-order theory without loss of generality. In
this paper, we will consider the Newtonian limit of the combined
Lagrangian
(\ref{eq:generals:Ls:0})-(\ref{eq:generals:Ls:1})-(\ref{eq:generals:Ls:2}),
see Sec.\ref{sec:feqs:general}, as a straightforward
generalization of the Einstein theory which is obviously recovered
in low curvature regimes.

As we said, several works focused on the cosmological implications
of additional terms to the Hilbert - Einstein Lagrangian
\cite{Capozziello:2002,Nojiri,Vollick:2003,Carroll:2004}. In
particular, terms of the form $R^{-n}$ (with positive $n$) have
been taken into account to explain the observed accelerated
behavior of the Hubble flow. Although such a term may lead to an
alternative explanation of the acceleration effect, its singular
behavior clearly leads to problems in the low curvature regime.
The explicit exclusion of flat solutions is in contradiction with
the basic assumptions of most of the weak field approximation
schemes of General Relativity, the most prominent examples being
the post-Minkowskian and the post-Newtonian approximation
\cite{Newton:2007}. In addition the non-validity of flat
solutions\footnote{As explicitly demanded by some authors, see p.\
3 in \cite{Carroll:2005}.} leads to the paradoxical situation that
standard linearization  procedures can no longer be based on
Minkowskian space in the lowest order and are therefore not well
defined in such a framework. While in a purely cosmological
context this drawback could be circumvented by basing the analysis
on a curved background, in the lowest order, it is not obvious how
such a theory could make sense at local scales\footnote{We use the
term local here for the rather broad range $\sim
10^{-2}-10^{11}$m, hence encompassing laboratory as well as
solar-system experiments.}.  In addition, it is not clear how one
could have a smooth and well-defined transition to General
Relativity for this kind of theories. Due to these drawbacks, we
will consider here theories which allow to recover the flat
solution.

The plan of the paper is the following. Sec.\ref{sec:feqs:general}
is devoted to give the general form of the fourth-order field
equations and their approximations at the lowest order in the weak
field limit, i.e. the Newtonian one. Sec.\ref{sec:generalapproach}
is devoted to the solutions of the field equations in the
Newtonian limit. In Sec.\ref{sec:green}, we develop in details the
method of Green functions for system with spherical symmetry,
while in Sec.\ref{sec:solgreen}, we exhibit explicit solutions
derived using the Green function. We draw conclusion in
Sec.\ref{sec:conclusions:outlook} and present a possible outlook
for future developments.  In \ref{app:approach}, we present an
alternative approach to solve the field equations, while  we
summarize  the conventions and the dimensions of quantities used
throughout the text in \ref{app:conventions}.

\section{Gravity with quadratic Lagrangians: The field equations and the Newtonian limit}
\label{sec:feqs:general}

In this Section, we discuss the  fourth-order field equations and
their Newtonian limit. This higher-order, with respect to the
standard second-order of Einstein field equations is due, as well
known,  to the integration of the boundary terms. These terms
disappear in General Relativity, thanks to the Divergence Theorem,
but this is not possible for several alternative theories of
gravity, as the higher-order ones, and then the derivative order
of field equations results augmented.

In this paper, we are interested to achieve the correct Newtonian
limit of gravity theories with quadratic Lagrangians in the
curvature invariants. This result can be achieved under two main
hypotheses: $i)$ asking for low velocities with respect to the
light speed and $ii)$ asking for week fields. By these requests,
the metric tensor is independent of time and second order
perturbation terms can be discarded in the field equations (see
also \cite{Newton:2007} for details). It is worth stressing that
the Newtonian limit of any relativistic theory of gravity  is
related to such hypotheses and it is a misunderstanding to
consider only the recovering of the Newtonian potential. In other
words,  a more general theory of gravity gives rise, in the
Newtonian limit, to gravitational potentials which can be  very
different from the standard Newtonian one.

In the third and in the  fourth parts of this Section,  we shall
discuss the field equations in the Newtonian limit  to show that
the fourth-order contributions to the potential  cannot be
trivially discarded.

\subsection{General form of the field equations}

Let us now come back to the choices displayed in
(\ref{eq:generals:Ls:0})-(\ref{eq:generals:Ls:1})-(\ref{eq:generals:Ls:2})-(\ref{eq:generals:Ls:3}),
for which the left-hand side of the field equations takes the
general form
%\footnote{Note the change of the sign of the first
%term in (\ref{eq:generals:gs:3}), compared to references
%\cite{DeWitt:1965,Havas:1977}, which comes from our convention for
%$R_{\alpha \beta \mu \nu}$.} :
\begin{eqnarray}
^{0}H_{\mu \nu } &=&R_{\mu \nu }-\frac{1}{2}g_{\mu \nu }R, \label{eq:generals:gs:0}\\
^{1}H_{\mu \nu } &=&2RR_{\mu \nu }-\frac{1}{2}g_{\mu \nu }R^{2}-2R_{;\mu \nu}+2g_{\mu \nu }\Box R, \label{eq:generals:gs:1}\\
^{2}H_{\mu \nu } &=&2R_{\mu}{}^{\alpha }R_{\nu \alpha
}-\frac{1}{2}g_{\mu \nu }R_{\alpha \beta}R^{\alpha \beta
}-2R_{(\mu |}{}^{\alpha }{}_{;|\nu )\alpha }+\Box R_{\mu\nu}\nonumber \\ && +g_{\mu \nu }R^{\alpha \beta }{}_{;\alpha \beta }, \label{eq:generals:gs:2} \\
^{3}H_{\mu \nu } &=&2R_{\mu \alpha \beta \gamma }R_{\nu
}{}^{\alpha \beta \gamma }-\frac{1}{2}g_{\mu \nu }R_{\alpha \beta
\gamma \delta }R^{\alpha \beta \gamma \delta}+4R_{\mu }{}^{\alpha
}{}_{\nu }{}^{\beta }\,_{;(\alpha \beta )}.
\label{eq:generals:gs:3}
\end{eqnarray}
All of the three expressions in
(\ref{eq:generals:gs:1})-(\ref{eq:generals:gs:2})-(\ref{eq:generals:gs:3})
involve fourth order differential operators. Due to the identity
\begin{eqnarray}
^{1}H_{\mu \nu } - 4 \,{}^{2}H_{\mu \nu } + \,{}^{3}H_{\mu \nu }
=0, \label{eq:identity:curvature}
\end{eqnarray}
which holds in a four-dimensional spacetime \cite{Lanczos:1938},
only two of the expressions in
(\ref{eq:generals:gs:1})-(\ref{eq:generals:gs:2})-(\ref{eq:generals:gs:3})
are independent, and we are free to use any two independent linear
combinations in our analysis. This identity gives rise to the well
known Gauss - Bonnet topological invariant which recently acquired
a lot of importance in cosmology as a possible source of dark
energy \cite{gaussbonnet}.  Furthermore, for a Lagrangian
comprising a general function of the Ricci scalar, we have
\begin{eqnarray}
{}^{f\left(R \right)}H_{\mu \nu}=\frac{df}{dR}R_{\mu \nu
}-\frac{1}{2}g_{\mu \nu}f-\biggl(\frac{df}{dR}\biggr)_{;\mu \nu
}+g_{\mu \nu }\Box\frac{df}{dR}\,.
\label{eq:feqs:fR:case:general}\end{eqnarray} Here we denoted the
covariant derivatives by a semicolon\footnote{We denote partial
derivatives with respect to the coordinates by a comma.} and the
d'Alembert operator by
$\Box$\footnote{$\Box\,=\,\frac{\partial_\alpha(\sqrt{-g}g^{\alpha\beta}\partial_\beta)}{\sqrt{-g}}$.}.
With these considerations in mind, let us to consider the
Newtonian limit of such a theory of gravity.

\subsection{The Newtonian limit}\label{sec:newtonian:limit}

Here we are not interested in entering the theoretical  discussion
on how to formulate a mathematically well sound Newtonian limit of
general relativistic field theories, for this we point the
interested reader to
\cite{Friedrichs:1927,Trautman:1963,Kilmister:1963,Dautcourt:1964,Kuenzle:1976,Ehlers:1980,Ehlers:1981}.
In this section, we provide the explicit form of the field
equations for the different admissible choices of Lagrangians
collected in Introduction at the lowest, i.e.\ Newtonian, order.
In the language of the post-Newtonian approximation, we are going
to consider the field equations up to the order ${\mathcal
O}\left(c^{-2} \right)$, where $c$ denotes the speed of light. For
the verification of our calculations we made use of a modified
version of the {\it Procrustes} package \cite{Puetzfeld:2006}.

We only mention, in passing, that there has also  been a
discussion of a somewhat alternative way to define the Newtonian
limit in higher-order theories in the recent literature, see for
example \cite{Dick:2004}. In this work, the Newtonian limit is
identified with the maximal symmetric solution, which is not
necessarily Minkowski spacetime in $f(R)$ theories which could be
singular.

Let us start from a flat background and work out the corresponding
field equations and hydrodynamic equations to the Newtonian order.
Our conventions are that $g_{\alpha \beta }$, with $\alpha ,\beta
=0,1,2,3$, can be transformed to $\eta _{\alpha \beta }=$
diag$\left( 1,-1,-1,-1\right) $ along a given curve. Latin indices
$i,j$ run from $1,2,3$, the coordinates are labelled by
$x^{\alpha}=\left( x^{0},x^{1},x^{2},x^{3}\right) =\left(
ct,x^{1},x^{2},x^{3}\right)$. We start with the following ansatz
for the metric
\begin{eqnarray}
g_{00} &=&1-\frac{2U}{c^{2}}+\mathcal{O}(c^{-4}),  \nonumber \\
g_{0a} &=&\frac{1}{c^3}h_{0\alpha}+\mathcal{O}(c^{-5}),  \nonumber \\
g_{ab} &=&-\left(1+\frac{2V}{c^{2}}\right)
\delta_{ab}+\mathcal{O}(c^{-4}).  \label{eq:covar:metric:cas}
\end{eqnarray}
%and try to determine the explicit form of $\sigma $, $h_{0a}$, and $V$ to the first post-Newtonian order\footnote{Note that we introduced $c$
%explicitly in the %time-space component, in this point we deviate from the notation used in \citet{Chandrasekhar:1965:2}. The reason to do this is
%linked to our use of computer algebra %methods, in which the chosen notation turns out to be more transparent.}. All of these components can depend
%on the spacetime coordinates, i.e.\ $U=U\left(x^{\alpha %}\right)$ etc., and $a=a(x^0)$ is a function of time only.
%With the metric ansatz from (\ref{eq:covar:metric:cas}) the line element takes the form
%\begin{eqnarray}
%ds^{2}&=&g_{\alpha \beta }dx^{\alpha }dx^{\beta } \nonumber\\
%     &=&\left( 1-\frac{2U}{c^{2}}+\frac{\sigma }{c^{4}}\right) dx^{0}dx^{0}+\frac{2h_{0a}}{c^3}dx^{0}dx^{a}\nonumber\\
%      &&-\left( 1+\frac{2V}{c^{2}}\right) \delta _{ab}dx^{a}dx^{b}. \label{eq:line:element}
%\end{eqnarray}
Apparently, the orders involved in this ansatz for the line
element reach beyond the Newtonian order, which we are mainly
interested in this work.
%The metric (\ref{eq:line:element}) allows
%us to give some outlook on the first post-Newtonian order within
%the theories under consideration.

On the matter side, i.e.\ right-hand side of the field equations,
we start with the general definition of the energy-momentum tensor
of a perfect fluid
\begin{eqnarray}
T_{\alpha \beta }=\left( \rho c^{2}+\Pi \rho +p\right) u_{\alpha }u_{\beta}-pg_{\alpha \beta },  \label{eq:emtensor:fluid}
\end{eqnarray}
here $\Pi$ denotes the internal energy  density, $\rho$ the energy
density, and $p$ the pressure. Following the procedure outlined in
\cite{Puetzfeld:2006:1}, we derive the explicit form of the
energy-momentum as follows
\begin{eqnarray}
T_{00}&=&\rho c^{2}+{\mathcal O}\left(c^{-2}\right),\label{eq:emtensor:00:cas}\\
T_{0a}&=&c\rho v^{a}+{\mathcal O}\left(c^{-1}\right),\label{eq:emtensor:0a:cas} \\
T_{ab}&=&\rho v^{a}v^{b}+p\delta_{ab}+{\mathcal
O}\left(c^{-2}\right)\label{eq:emtensor:ab:cas}.
\end{eqnarray}
The general form of the field equations is given by
\begin{eqnarray}
H_{\mu \nu}\,=\,\frac{8 \pi G}{c^4} T_{\mu \nu},
\label{eq:feqs:general}
\end{eqnarray}
% Old sign
%\begin{eqnarray}
%G_{\mu \nu}=\kappa T_{\mu \nu}=-\frac{8 \pi G}{c^4} T_{\mu \nu}, \label{eq:feqs:general}
%\end{eqnarray}
with a generalized  tensor $H_{\mu \nu}$, being a combination of
the expressions\footnote{Obviously ${}^0H_{\mu\nu}$ is the
Einstein tensor.} specified in
(\ref{eq:generals:gs:0})-(\ref{eq:generals:gs:1})-(\ref{eq:generals:gs:2})-(\ref{eq:generals:gs:3}),
which, in turn, depend on the final form of the Lagrangian.

\subsection{The quadratic Lagrangians and the Newtonian limit of the field equations}\label{sec:feqs:general_aws}

Let us  consider now the field equations, in the Newtonian limit,
for the possible quadratic Lagrangians which we compare to the
Newtonian limit of the standard Hilbert - Einstein Lagrangian. It
is important to stress that the field equations, in Newtonian
limit, are considered up to the order $\mathcal{O}(c^{-2})$ while
the vector component to the order $\mathcal{O}(c^{-3})$ is related
to the post-Newtonian limit of the theory (see \cite{gravmagn} for
details). Up to the Newtonian order the left-hand side of the
field equations, i.e.\
(\ref{eq:generals:gs:0})-(\ref{eq:generals:gs:1})-(\ref{eq:generals:gs:2})-(\ref{eq:generals:gs:3}),
takes the following form\footnote{Here we made use of the
following operator definition
$\nabla^{2}:=\delta^{ab}\frac{\partial^{2}}{\partial x^{a}\partial
x^{b}}$, as well as $\nabla^{4}:=\nabla^{2} \nabla^{2}$.} for the
metric given in (\ref{eq:covar:metric:cas}):

\begin{itemize}
\item The Hilbert - Einstein Lagrangian ($\mathcal{L}_0\,=\,R$)
\begin{eqnarray}
^{0}H_{00} &=&- \frac{2}{c^{2}} \nabla^{2}V, \label{eq:0G:1}\\
^{0}H_{0a} &=&0, \label{eq:0G:2}\\
^{0}H_{ab}&=&\frac{1}{c^{2}}\left[\left(\nabla^{2}V-\nabla^{2}U\right)\delta_{ab}-\left(V-U\right)_{,ab}\right];\label{eq:0G:3}
\end{eqnarray}
\item The $R^2$ - Lagrangian ($\mathcal{L}_1\,=\,R^2$)
\begin{eqnarray}
^{1}H_{00} &=&\frac{4}{c^{2}}\left( 2\nabla ^{4}V-\nabla ^{4}U\right), \label{eq:1G:1}\\
^{1}H_{0a} &=&0, \label{eq:1G:2}\\
^{1}H_{ab}&=&\frac{4}{c^{2}}\left[\left(\nabla^{4}U-2\nabla^{4}V\right)\delta_{ab}+\left(2\nabla^{2}V-\nabla^{2}U\right)_{,ab}\right];\label{eq:1G:3}
\end{eqnarray}
\item The $R_{\alpha \beta} R^{\alpha \beta}$ - Lagrangian ($\mathcal{L}_2\,=\,R_{\alpha \beta} R^{\alpha \beta}$)
\begin{eqnarray}
^{2}H_{00} &=&\frac{2}{c^{2}}\left( \nabla ^{4}V-\nabla ^{4}U\right) , \label{eq:2G:1} \\
^{2}H_{0a} &=&0 , \label{eq:2G:2}\\
^{2}H_{ab} &=&\frac{1}{c^{2}}\left[ \left( \nabla ^{4}U-3\nabla
^{4}V\right)\delta _{ab}+\left( 3\nabla ^{2}V-\nabla ^{2}U\right)
_{,ab}\right]; \label{eq:2G:3}
\end{eqnarray}
\item The $R_{\alpha \beta \gamma \delta} R^{\alpha \beta \delta \gamma}$ - Lagrangian ($\mathcal{L}_3\,=\,R_{\alpha \beta \gamma \delta} R^{\alpha \beta
\delta \gamma}$)
\begin{eqnarray}
^{3}H_{00} &=&-\frac{4}{c^{2}}\nabla ^{4}U , \label{eq:3G:1} \\
^{3}H_{0a} &=&0 , \label{eq:3G:2} \\
^{3}H_{ab} &=&-\frac{4}{c^{2}}\left( \nabla ^{4}V\delta
_{ab}-\nabla^{2}V_{,ab}\right). \label{eq:3G:3}
\end{eqnarray}
\end{itemize}

\subsection{The combined Lagrangian}\label{subsec:choice:combined}

Let us now combine the different terms of the last section into
the same Lagrangian. This combination is the basis for our
investigation for the remaining part of the paper. Since terms
resulting from $R^n$ with $n \geq 3$ do {\it not} contribute at
the order $\mathcal{O}\left( c^{-2} \right)$ , the most general
choice for the Lagrangian is
\begin{eqnarray}
\mathcal{L}=a_1 R + a_2 R^2 + b_1 R_{\mu \nu }R^{\mu \nu } + c_1
R_{\alpha \beta \mu \nu }R^{\alpha \beta \mu \nu}
\label{eq:lag:most:general:quad:1}.
\end{eqnarray} Due to the identity given in (\ref{eq:identity:curvature}), it is
sufficient to study
\begin{eqnarray}
\mathcal{L}=a_1 R + a_2 R^2 + b_1 R_{\mu \nu }R^{\mu \nu }
\label{eq:lag:most:general:quad:2},
\end{eqnarray}
in four dimension, where we introduced the constants $a_1$, $a_2$,
$b_1$\footnote{Note that $[a_1]={[\rm length]}^{0}$,
$[a_2]=[b_1]={[\rm length]}^{2}$.}\footnote{The coefficients of
(\ref{eq:lag:most:general:quad:1}) are different from ones of
(\ref{eq:lag:most:general:quad:2}).}. If we take into account the
results from (\ref{eq:emtensor:00:cas})-(\ref{eq:emtensor:ab:cas})
as well as (\ref{eq:0G:1})-(\ref{eq:2G:3}) the explicit form of
the field equations (\ref{eq:feqs:general}) up to the Newtonian
order is

\begin{eqnarray}\label{fieldeq1}\label{eq:combined:1}
-2a_1\nabla^2V+(8a_2+2b_1)\nabla^4V-(4a_2+2b_1)\nabla^4U\,=\,8\pi G \rho\,,\\\nonumber\\
\biggl[a_1(\nabla^2V-\nabla^2U)-(8a_2+3b_1)\nabla^4V+(4a_2+b_1)\nabla^4U\biggr]\delta_{ab}
\nonumber\\+\biggl[(8a_2+3b_1)\nabla^2V-(4a_2+b_1)\nabla^2U+a_1(U-V)\biggr]_{,ab}\,=\,0\label{fieldeq2}\label{eq:combined:3}
\end{eqnarray}
the equations which we are going to solve.

\section{Considerations on the field equations in the Newtonian limit}\label{sec:generalapproach}

 In this Section, we are going to formulate the problem to solve
the field equations   (\ref{fieldeq1}) - (\ref{fieldeq2}) in the
most general way. It is worth noticing that   the isotropic
coordinates for the metric (\ref{eq:covar:metric:cas}) allow to
search for solutions independently of the  symmetry of the
physical system (which can be spherical, cylindrical etc.). The
results which we are going to achieve are completely general since
we will search for solutions in terms of Green functions. However,
being the combined Lagrangian (\ref{eq:lag:most:general:quad:2})
built up by various terms, the field equations strictly depend on
the coupling constants. As we will see, the value of such
coefficients have a crucial role for the validity of the approach
since, from a physical viewpoint, we have to obtain the Newtonian
limit of General Relativity as soon as the quadratic corrections
disappear. This aspect of the problem is not accurately faced in
the literature and can lead to wrong conclusions. Our aim is to
develop a method which allows to control, step by step, the
Newtonian limit in agreement with the results of General
Relativity. This is possible at three levels: Lagrangian, field
equations and solutions. In the second part of this Section, we
will analyze the various cases of field equations considering
particular values of the coefficients. Specifically, we will take
into account the values where the proposed approach fails. It is
interesting to note that any time  the Hilbert - Einstein term is
absent into the Lagrangian, the field equations are fourth order
(see Table \ref{tablefieldequation}). In other words, if the
Hilbert - Einstein term is not present, we do not recover the
Laplace/Poisson equations.

\subsection{The general approach to decouple the field
equations}\label{general_approach}

By introducing two new auxiliary functions ($A$ and $B$), the
equations (\ref{eq:combined:1})-(\ref{eq:combined:3}) become

\begin{eqnarray}\label{eq3}&\nabla^2\biggl\{\frac{4a_2+b_1}{2a_2+b_1}A+\frac{a_1}{2a_2+b_1}B+\nabla^2\biggr[\frac{2b_1
(3a_2+b_1)} {a_1(2a_2+b_1)}A\nonumber\\&
-\frac{2a_2}{2a_2+b_1}B\biggr]\biggr\}=8\pi G\rho\,,\end{eqnarray}

\begin{equation}\label{eq4}\nabla^2(A+\nabla^2B)\delta_{ab}-(A+\nabla^2B)_{,ab}=0\,,\end{equation}
where $A$ and $B$ are linked to $U$ and $V$ via

\begin{equation}\label{trans1}A:=a_1(V-U)\,,\end{equation}

\begin{equation}\label{trans2}B:=(4a_2+b_1)V-(8a_2+3b_1)U\,.\end{equation}
%The inverse transformations are given by
%\begin{equation}\label{pot1}U=-\frac{(8a_2+3b_1)A+a_1B}{2a_1(2a_2+b_1)}\,,\end{equation}
%\begin{equation}\label{pot2}V=-\frac{(4a_2+b_1)A+a_1B}{2a_1(2a_2+b_1)}\,.\end{equation}
Obviously we must require $a_1(2a_2+b_1)\neq 0$, which is the
determinant of the transformations (\ref{trans1})-(\ref{trans2}).
Let us introduce the new function $\Phi$ defined as follows:

\begin{equation}\label{definition}\Phi:=A+\nabla^2B.\end{equation}
At this point, we can use the new function $\Phi$ to decouple the
system (\ref{eq3})-(\ref{eq4}). In fact we obtain

\begin{eqnarray}\label{eq5}&-\frac{2b_1(3a_2+b_1)}
{a_1(2a_2+b_1)}\nabla^6B-\frac{6a_2+b_1}{2a_2+b_1}\nabla^4B+\frac{a_1}{2a_2+b_1}\nabla^2B\nonumber\\&=
8\pi G\rho-\nabla^2\tau_I\,,&\end{eqnarray}

\begin{equation}\label{eq6}\nabla^2\Phi\delta_{ab}-\Phi_{,ab}=0\,,\end{equation}
where ${\displaystyle
\tau_I:=\frac{4a_2+b_1}{2a_2+b_1}\Phi+\frac{2b_1(3a_2+b_1)}
{a_1(2a_2+b_1)}\nabla^2\Phi}$. We are interested in the solution
of (\ref{eq5}) in terms of the Green function
$\mathcal{G}_I(\mathbf{x},\mathbf{x}')$ defined by

\begin{equation}\label{gr1}B(\mathbf{x})=Y_I\int d^3\mathbf{x}'
\mathcal{G}_I(\mathbf{x},\mathbf{x}')\sigma_I(\mathbf{x}')\,,\end{equation}
where
\begin{equation}\label{den1}\sigma_I(\mathbf{x}):=8\pi
G\rho(\mathbf{x})-\nabla^2\tau_I(\mathbf{x})\,,\end{equation} and
$Y_I$ being a constant, which we introduced for dimensional
reasons. Then the set of equations
(\ref{eq:combined:1})-(\ref{eq:combined:3}) is equivalent to

\begin{eqnarray}\label{sys11}
&\frac{2b_1(3a_2+b_1)}
{a_1(2a_2+b_1)}\nabla^6_{\mathbf{x}}\mathcal{G}_I(\mathbf{x},\mathbf{x}')+\frac{6a_2+b_1}{2a_2+b_1}\nabla^4_{\mathbf{x}}\mathcal{G}_I(\mathbf{x},
\mathbf{x}')\nonumber\\&-\frac{a_1}{2a_2+b_1}\nabla^2_{\mathbf{x}}\mathcal{G}_I(\mathbf{x},\mathbf{x}')=
-Y_I^{-1}\delta(\mathbf{x}-\mathbf{x}')\,,\end{eqnarray}

\begin{equation}\label{sys12}\nabla^2\Phi(\mathbf{x})\delta_{ab}-\Phi(\mathbf{x})_{,ab}=0\,,
\end{equation}
where $\delta(\mathbf{x}-\mathbf{x}')$ is the 3-dimensional Dirac
$\delta$-function. The general solutions of equations
(\ref{eq:combined:1})-(\ref{eq:combined:3}) for $U(\mathbf{x})$
and $V(\mathbf{x}$), in terms of the Green function
$\mathcal{G}_I(\mathbf{x}, \mathbf{x}')$ and the function
$\Phi(\mathbf{x})$, are

\begin{eqnarray}\label{sol5}U(\mathbf{x})&=&Y_I\frac{(8a_2+3b_1)\nabla_\mathbf{x}^2-a_1}{2a_1(2a_2+b_1)}\int
d^3\mathbf{x}'\mathcal{G}_I(\mathbf{x},\mathbf{x}')\biggl[8\pi
G\rho(\mathbf{x}')\nonumber\\&&-\frac{4a_2+b_1}{2a_2+b_1}\nabla^2_{\mathbf{x}'}\Phi(\mathbf{x}')-\frac{2b_1(3a_2+b_1)}{a_1(2a_2+b_1)}\nabla_{\mathbf{x}'}
^4\Phi(\mathbf{x}')\biggr]\nonumber\\&&
-\frac{8a_2+3b_1}{2a_1(2a_2+b_1)}\Phi(\mathbf{x})\,,
\end{eqnarray}

\begin{eqnarray}\label{sol6}V(\mathbf{x})&=&Y_I\frac{(4a_2+b_1)\nabla_\mathbf{x}^2-a_1}{2a_1(2a_2+b_1)}\int
d^3\mathbf{x}'\mathcal{G}_I(\mathbf{x},\mathbf{x}')\biggl[8\pi
G\rho(\mathbf{x}')\nonumber\\&&-\frac{4a_2+b_1}{2a_2+b_1}\nabla^2_{\mathbf{x}'}\Phi(\mathbf{x}')-\frac{2b_1(3a_2+b_1)}{a_1(2a_2+b_1)}\nabla_{\mathbf{x}'}
^4\Phi(\mathbf{x}')\biggr]\nonumber\\&&-\frac{4a_2+b_1}{2a_1(2a_2+b_1)}\Phi(\mathbf{x})\,.\end{eqnarray}
Equations (\ref{eq:combined:1}) - (\ref{eq:combined:3}) represent
a coupled set of fourth order differential equations. The total
number of integration constants is eight. With the substitution
(\ref{definition}), it has been possible to decouple the set of
equations, but now the differential order is changed. The total
differential order is the same, indeed we have one equation of
sixth order (\ref{eq5}), and another equation of second order
(\ref{eq6}), while previously we had two equations of fourth
order. Obviously, the number of integration constants is
conserved.   The possibility to decouple  the field equations
(\ref{fieldeq1}) - (\ref{fieldeq2}) is strictly realted to the
choice to express the auxiliary field  $A$ in terms of $B$  by
inverting the relation (\ref{definition}), deriving Eq.
(\ref{fieldeq1}) and reducing (\ref{fieldeq2}) to the second
order. In \ref{app:approach}, we discuss a different where the set
of equations (\ref{fieldeq1}) - (\ref{fieldeq2}) remains of forth
order by using the relation between $B$ and $A$.

\subsection{Field equations for  particular values of the coupling
constants}\label{particularfieldequation}

In this subsection, we want to analyze the behavior of the field
equations  (\ref{fieldeq1}) - (\ref{fieldeq2}) for those values of
the coupling constants $a_1$, $a_2$, $b_1$ where the
transformations (\ref{trans1}) - (\ref{trans2}) do not hold.
Specifically, in Table \ref{tablefieldequation}, we display
several  cases of the field equations (\ref{eq:combined:1}) -
(\ref{eq:combined:3}), for different choices of the coupling
constants, where the determinant of transformations (\ref{trans1})
- (\ref{trans2}) is zero. First of all, we have to note that, for
$a_2\,=\,b_1\,=\,0$ (Case i in the Table
\ref{tablefieldequation}), we trivially obtain the  same result of
General Relativity in isotropic coordinates. Furthermore, by
asking for $U\,=\,V$, the spatial equation is satisfied. It is
straightforward to derive as solution the Newton potential  (see
$\S$ \ref{particularsolution} for details). Particularly
interesting is also  Case iv,  where both terms
$R_{\alpha\beta}R^{\alpha\beta}$ and $R^2$ give  similar
contributions. In other words, we have the same situation of the
Lagrangian  ${\cal L}=a_1R+a_2R_{\alpha\beta}R^{\alpha\beta}$ with
a redefinition of the couplings. However, this Lagrangian is
compatible with the transformations (\ref{trans1}) -
(\ref{trans2}) and then the results of $\S$ \ref{general_approach}
hold. In Cases  (ii - iii - v - vi - vii),  the differential
operator $\nabla^2$  never appears as a linear term since the
invariants $R^2$ and $R_{\alpha\beta}R^{\alpha\beta}$  give rise
to higher order terms in the field equations. In these cases, the
full field equations (not in the weak field regime) give
$g_{\mu\nu}\Box R-R_{;\mu\nu}$ for the Lagrangian $R^2$ and
$-2R_{(\mu |}{}^{\alpha }{}_{;|\nu )\alpha }+\Box
R_{\mu\nu}+g_{\mu \nu }R^{\alpha \beta }{}_{;\alpha \beta }$ for
$R_{\alpha\beta}R^{\alpha\beta}$ which are fourth order equations.
In the weak field regime, one obtains  the equations  reported in
Table \ref{tablefieldequation}. As we will see in $\S$
\ref{particularsolution} devoted to the solutions, all these cases
do not present a Newtonian potential.

\begin{table}[ht]
\centering
\begin{tabular}{c|c|c}
\hline\hline\hline
 Cases & Choices of $a_1$, $a_2$, $b_1$ & Corresponding field equations \\
 \hline
 & & \\
 i & $\begin{array}{ll}a_2=0\\b_1=0\end{array}$ & $\begin{array}{ll}
 \nabla^2V=-\frac{4\pi G}{a_1}\rho\,,\\\nabla^2\biggl[V-U\biggr]\delta_{ab}-\biggl[V-U\biggr]_{,ab}=0\end{array}$ \\
 \hline
 & & \\
 ii & $\begin{array}{ll}a_1=0\\b_1=0\end{array}$ & $\begin{array}{ll}\nabla^4(2V-U)=\frac{2\pi G}{a_2}\rho\,,\\\nabla^2\biggl[\nabla^2(2V-U)\biggr]
 \delta_{ab}-
 \biggl[\nabla^2(2V-U)\biggr]_{,ab}=0
 \end{array}$ \\
 \hline
 & & \\
 iii & $\begin{array}{ll}a_1=0\\a_2=0\end{array}$ &
 $\begin{array}{ll}\nabla^4(U-V)=-\frac{4\pi
 G}{b_1}\rho\,,\\\nabla^2\biggl[\nabla^2
 (U-3V)\biggr]\delta_{ab}-\biggl[\nabla^2(U-3V)\biggr]_{,ab}=0\end{array}$ \\
 \hline
 & & \\
 iv & $b_1=-2a_2$ &
 $\begin{array}{ll}2a_2\nabla^4V-a_1\nabla^2V=4\pi G\rho\,,\\\nabla^2\biggl[a_1(V-U)-2a_2\nabla^2(V-U)\biggr]\delta_{ab}\\-\biggl[a_1(V-U)-2a_2\nabla^2(V
 -U)\biggr]_{,ab}=0\end{array}$ \\
 \hline
 & & \\
 v & $\begin{array}{ll}a_1= 0\\b_1=-4a_2\end{array}$ & $\begin{array}{ll}\nabla^4U=\frac{2\pi G}{a_2}\rho\,,\\\nabla^2\biggl[\nabla^2V\biggr]\delta_{ab}-
 \biggl[\nabla^2V\biggr]_{,ab}=0\end{array}$ \\
 \hline
 & & \\
 vi & $\begin{array}{ll}a_1=0\\b_1=-2a_2\end{array}$ & $\begin{array}{ll}\nabla^4V=\frac{2\pi G}{a_2}\rho\,,\\\nabla^2\biggl[\nabla^2(V-U)\biggr]\delta_
 {ab}-\biggl[\nabla^2(V-U)\biggr]_{,ab}=0\,;\end{array}$ \\
 \hline
 & & \\
 vii & $\begin{array}{ll}a_1=0\\b_1=-\frac{8a_2}{3}\end{array}$ & $\begin{array}{ll}\nabla^4(2V+U)=\frac{6\pi G}{a_2} \rho\,,\\\nabla^2\biggl[\nabla^2U
 \biggr]\delta_{ab}-
 \biggl[\nabla^2U\biggr]_{,ab}=0\end{array}$ \\
 \hline\hline\hline
 \end{tabular}
\caption{\label{tablefieldequation}Explicit form of the field
equations for different choices of the coupling constants for
which the determinant of the transformations
(\ref{trans1})-(\ref{trans2}) vanishes. Cases i, ii, iii are the
Lagrangians introduced in the Sec.\ref{sec:feqs:general_aws} ($R$,
$R^2$, $R_{\mu\nu}R^{\mu\nu}$).}
\end{table}

\section{Green's functions for  spherically
symmetric systems}\label{sec:green}

We are interested in the solutions of field equations
(\ref{eq:feqs:general}) at order ${\mathcal O}(c^{-2})$ by using
the method of Green functions.  We have to stress that the method
of Green's functions does not work in the general case since the
field equations are non-linear. However, the Newtonian limit of
the theory  (based on the hypothesis that metric nonlinear terms
can be discarded) allows that also the field equations result
linearized). By solving the field equations with the Green
function method, one obtains, as a first result, the solution in
terms of gravitational potential in the point-mass case. Then by
using  the equations (\ref{sol5})-(\ref{sol6}), obtained
in the weak field limit and then in Newtonian linear approximation
 for a spatial distribution of matter, we obtain, in principle,
the gravitational potential for a given density profile. If the
matter possesses  a spherical symmetry, also the Green function
has to be spherically symmetric. In this case, the correlation
between two points has to be a function of the radial coordinate
only, that is:
$\mathcal{G}(\mathbf{x},\mathbf{x}')=\mathcal{G}(|\mathbf{x}-\mathbf{x}'|)$.
It is important to stress again the fact that the approach
works if and only if we are in the linear approximation, i.e. in
the Newtonian limit.

\subsection{A general Green function for the decoupled field equations}\label{sec:greenfunctparagraph}

Let us introduce the radial coordinate
$r:=|\mathbf{x}-\mathbf{x}'|$; with this choice, equation
(\ref{sys11}) for $r\neq 0$ becomes

\begin{equation}\label{eq16}2b_1(3a_2+b_1)\nabla_r^6\mathcal{G}_I(r)+a_1(6a_2+b_1)\nabla_r^4\mathcal{G}_I(r)-a^2_1\nabla_r^2
\mathcal{G}_I(r)=0\,,\end{equation} where
$\nabla^2_r=r^{-2}\partial_r(r^{-2}\partial_r)$ is the radial
component of the Laplacian in polar coordinates. The solution of
(\ref{eq16}) is:

\begin{eqnarray}\label{sol9}\mathcal{G}_I(r)&=&K_{I,1}-\frac{1}{r}\biggr[K_{I,2}+\frac{b_1}{a_1}\biggl(K_{I,3}e^{-\sqrt{-\frac{a_1}{b_1}}r}
+K_{I,4}e^{\sqrt{-\frac{a_1}{b_1}}r}\biggr)\nonumber\\&&-\frac{2(3a_2+b_1)}{a_1}\biggl(K_{I,5}e^{-\sqrt{\frac{a_1}{2(3a_2+b_1)}}r}
+K_{I,6}e^{\sqrt{\frac{a_1}{2(3a_2+b_1)}}r}\biggr)\biggr]
\end{eqnarray}
where $K_{I,1}$, $K_{I,2}$, $K_{I,3}$, $K_{I,4}$, $K_{I,5}$,
$K_{I,6}$ are constants. The integration constants $K_{I,i}$ have
to be fixed by imposing the boundary conditions at infinity and in
the origin. A physically acceptable solution has to satisfy the
condition $\mathcal{G}(\mathbf{x},\mathbf{x}')\rightarrow 0$ if
$|\mathbf{x}-\mathbf{x}'|\rightarrow \infty$, then the constants
$K_{I,1}$, $K_{I,4}$, $K_{I,6}$ in equation (\ref{sol9}) have to
vanish. We note that, if $a_2=b_1=0$, the Green function of the
Newtonian mechanics is found. In this case, we have the complete
analogy with the Electromagnetism.  More precisely, when we do not
consider higher-order tems than  Hilbert - Einstein one in the
gravitational Lagrangian, we obtain, in the Newtonian limit, a
field equation analog to the electromagnetic one for the scalar
component (electric potential). This means that we have the same
form of the Green function \cite{jackson}.

To obtain the conditions on the constants $K_{I,2}$, $K_{I,3}$,
$K_{I,5}$ we consider the Fourier transform of
$\mathcal{G}(\mathbf{x},\mathbf{x}')$:

\begin{equation}\mathcal{G}_I(\mathbf{x},\mathbf{x}')=\int\frac{d^3\mathbf{k}}{(2\pi)^{3/2}}\,\,\tilde{\mathcal{G}}_I
(\mathbf{k})\,\,e^{i\mathbf{k}\cdot(\mathbf{x}-\mathbf{x}')}\,.\end{equation}
$\mathcal{G}_I(\mathbf{x},\mathbf{x}')$ depends on the nature of
the poles of $|\mathbf{k}|$ and on the values of the arbitrary
constants $a_1$, $a_2$, $b_1$. If we define two new quantities
$\lambda_1$, $\lambda_2 \in \,\mathcal{R}$:

\begin{equation}\lambda_1^2:=-\frac{a_1}{b_1}\,,\,\,\,\,\,\,\,\,\,\,\,\,\,\,\,\,\,\lambda_2^2:=\frac{a_1}{2(3a_2+b_1)}\,,
\label{scale2}\end{equation} we obtain:

\begin{equation}\label{green_function}\mathcal{G}_I(\mathbf{x},\mathbf{x}')=\sqrt{\frac{\pi}{18}}\frac{Y^{-1}_I}{|\mathbf{x}-\mathbf{x}'|}\biggl[
\frac{\lambda_2^2-\lambda_1^2}{\lambda_1^2\lambda_2^2}-\frac{e^{-\lambda_1|\mathbf{x}-\mathbf{x}'|}}{\lambda_1^2}+\frac{e^{-\lambda_2|\mathbf{x}-
\mathbf{x}'|}}{\lambda_2^2}\biggr]\,.\end{equation} This Green
function corresponds to the one in (\ref{sol9}). Obviously, we
have three possibilities for the parameters $\lambda_1$ and
$\lambda_2$.  In fact, $\lambda_1$ and $\lambda_2$ are related to
the algebraic signs of $a_1$, $a_2$, $b_1$ and then we can also
achieve real values for such parameters. This means that we  have
three possibilities: both imaginary, one real and one imaginary.
In Table \ref{tablegrennfunction}, we provide the complete set of
Green functions $\mathcal{G}_{I}(\mathbf{x},\mathbf{x}')$,
depending on the choices of the coefficients $a_2$ and $b_1$ (with
a fixed sign of $a_1$). The various modalities in which we obtain
the Green functions are due to the various sign combinations of
the arbitrary constants. In general, the parameters
$\lambda_{1,2}$ indicate characteristic scale lengths where
corrections to the Newtonian potential can be appreciated. It is
worth noticing that, thanks to the forms of the Green functions
(see Table \ref{tablegrennfunction}),  the Newton behavior is
always asymptotically recovered.
\begin{table}[ht]
 \centering
 \begin{tabular}{c|c|c}
 \hline\hline\hline
 Cases & Choices of $a_2$, $b_1$ & Green function $\mathcal{G}_I(\mathbf{x},\mathbf{x}')$ \\
 \hline
 & & \\
 viii & $\begin{array}{ll}b_1<0\\\\3a_2+b_1>0\end{array}$ & $ \sqrt{\frac{\pi}{18}}\frac{Y^{-1}_I}{|\mathbf{x}-\mathbf
 {x}'|}\biggl[\frac{\lambda_2^2-\lambda_1^2}
{\lambda_1^2\lambda_2^2}-\frac{e^{-\lambda_1|\mathbf{x}-\mathbf{x}'|}}{\lambda_1^2}+\frac{e^{-\lambda_2|\mathbf{x}-\mathbf{x}'|}}{\lambda_2^2}\biggr]$ \\
 \hline
 & & \\
 ix & $\begin{array}{ll}b_1>0\\\\3a_2+b_1<0\end{array}$ &
 $ \sqrt{\frac{\pi}{18}}\frac{Y^{-1}_I}{|\mathbf{x}-\mathbf
 {x}'|}\biggl[\frac{\lambda_1^2-\lambda_2^2}{\lambda_1^2\lambda_2^2}+\frac{\cos(\lambda_1|\mathbf{x}-\mathbf{x}'|)}{\lambda_1^2}-\frac{\cos(\lambda_2|
 \mathbf{x}-\mathbf{x}'|)}{\lambda_2^2}\biggr]$ \\
 \hline
 & & \\
 x & $\begin{array}{ll}b_1<0\\\\3a_2+b_1<0\end{array}$ &
 $ \sqrt{\frac{\pi}{18}}\frac{Y^{-1}_I}{|\mathbf{x}-\mathbf{x}'|}\biggl[\frac{\lambda_1^2+\lambda_2^2}{\lambda_1^2
 \lambda_2^2}-\frac{e^{-\lambda_1|\mathbf{x}-\mathbf{x}'|}}{\lambda_1^2}-\frac{\cos(\lambda_2|\mathbf{x}-\mathbf{x}'|)}{\lambda_2^2}\biggr]$ \\
 \hline\hline\hline
 \end{tabular}
 \caption{\label{tablegrennfunction}The complete set of Green
 functions for equations (\ref{sys11}). The scale lengths are: $\lambda_1:=|a_1/b_1|^{1/2}$,
 $\lambda_2:=|a_1/2(3a_2+b_1)|^{1/2}$. It is possible to have a further choice for the scale lengths which turns out to be
 dependent on the two knows length scales. In fact, if we perform the substitution
 $\lambda_1\rightleftharpoons\lambda_2$, we obtain a fourth choice. In addition, for a correct
 Newtonian component, we assumed $a_1>0$. In fact when $a_2=b_1=0$
 the field equations (\ref{eq:combined:1}) and
 (\ref{eq:combined:3}) give us the Newtonian theory of gravity if $a_1=1$.}
\end{table}
When one considers a point-like source,
$\rho\propto\delta(\mathbf{x})$, and by setting
$\Phi(\mathbf{x})=0$ the potentials (\ref{sol5})-(\ref{sol6}) are
proportional to $\mathcal{G}_I(\mathbf{x},\mathbf{x}')$. Without
losing generality we have:

\begin{equation}\label{potpointsource}U(\mathbf{x})\sim\frac{U_0}{\mathbf{|x|}}+U_1\frac{e^{-\lambda_1\mathbf{|x|}}}{\mathbf{|x|}}+U_2\frac{e^{-\lambda_2
\mathbf{|x|}}}{\mathbf{|x|}}\,,\end{equation} where $U_0$, $U_1$,
$U_2$ are some integration constants. An analogous behavior is
obtained for the potential $V(\mathbf{x})$. We note that in the
vacuum case we found a Yukawa-like corrections to Newtonian
mechanics  but with two scale lengths related to the quadratic
corrections in the Lagrangian (\ref{eq:lag:most:general:quad:2})
(see also the above expressions (\ref{scale2})). This behavior is
strictly linked to the sixth order of (\ref{sys11}), which depends
on the coupled form of the system of equations
(\ref{eq:combined:1})-(\ref{eq:combined:3}). In fact if we
consider the Fourier transform of the potentials $U$ and $V$:

\begin{equation}U(\mathbf{x})=\int\frac{d^3\mathbf{k}}{(2\pi)^{3/2}}\,\,\tilde{u}(\mathbf{k})\,\,e^{i\mathbf{k}\cdot\mathbf{x}}\,,\,\,\,\,\,\,\,\,
V(\mathbf{x})=\int\frac{d^3\mathbf{k}}{(2\pi)^{3/2}}\,\,\tilde{v}(\mathbf{k})\,\,e^{i\mathbf{k}\cdot\mathbf{x}}\,,\end{equation}
the solutions of equations (\ref{fieldeq1})-(\ref{fieldeq2}) are

\begin{equation}\label{potfouU}U(\mathbf{x})=\int\frac{d^3\mathbf{k}}{(2\pi)^{3/2}}\frac{4\pi G[a_1+(8a_2+3b_1)\mathbf{k}^2]\tilde{\rho}
(\mathbf{k})e^{i\mathbf{k}\cdot\mathbf{x}}}{\mathbf{k}^2(a_1-b_1\mathbf{k}^2)[a_1+2(3a_2+b_1)\mathbf{k}^2]}\,,\end{equation}

\begin{equation}\label{potfouV}V(\mathbf{x})=\int\frac{d^3\mathbf{k}}{(2\pi)^{3/2}}\frac{4\pi G[a_1+(4a_2+b_1)\mathbf{k}^2]\tilde{\rho}
(\mathbf{k})e^{i\mathbf{k}\cdot\mathbf{x}}}{\mathbf{k}^2(a_1-b_1\mathbf{k}^2)[a_1+2(3a_2+b_1)\mathbf{k}^2]}\,,\end{equation}
where $\tilde{\rho}(\mathbf{k})$ is the Fourier transform of the
matter density.
%We can see that the solutions have the same poles
%as (\ref{intfou}).
It is possible to show, by applying the Fourier transform to the
potentials $U$ and $V$, that the poles in Eqs. (\ref{potfouU}) -
(\ref{potfouV}) are always three.

%It is worth noticing that any
%pole brings two differential orders. However, also with three
%poles, the system is fourth-order and not sixth-order, as it seems
%from Eq. (\ref{sys11}). In fact, the differential order results
%increased since the set of differential equations is coupled but,
%after the decoupling the correct order is restored.

Finally, if ${\displaystyle
\tilde{\rho}(\mathbf{k})=\frac{M}{(2\pi)^{3/2}}}$ (the Fourier
transform of a point-like source) the solutions
(\ref{potfouU})-(\ref{potfouV}) are similar to
(\ref{potpointsource}). In fact, if we suppose that $b_1\neq 0$
and $3a_2+b_1\neq0$, the solutions (\ref{potfouU}) -
(\ref{potfouV}) are

\begin{equation}\label{potfouU1}U(\mathbf{x})=\frac{GM}{a_1|\mathbf{x}|}\biggl(1-\frac{4}{3}e^{-\lambda_1|\mathbf{x}|}+\frac{1}{3}
e^{-\lambda_2|\mathbf{x}|}\biggr)\,,\end{equation}

\begin{equation}\label{potfouV1}V(\mathbf{x})=\frac{GM}{a_1|\mathbf{x}|}\biggl(1-\frac{2}{3}e^{-\lambda_1|\mathbf{x}|}-\frac{1}{3}
e^{-\lambda_2|\mathbf{x}|}\biggr)\,.\end{equation}

\subsection{Green functions for particular values of the coupling constants}

 The obtained Green functions deserve some comments. First of
all, we have to consider the particular  values of the parameters
where the general approach developed in $\S$
\ref{sec:greenfunctparagraph} does not work. For example,  if
$b_1=0$, we have only one Yukawa-like correction. The Green
function have to satisfy the equation

\begin{eqnarray}\label{greebequationfR}3\,\nabla^4_{\mathbf{x}}\mathcal{G}_I(\mathbf{x},\mathbf{x}')-\frac{a_1}
{2a_2}\nabla^2_{\mathbf{x}}\mathcal{G}_I(\mathbf{x},\mathbf{x}')=-Y_I^{-1}\delta(\mathbf{x}-\mathbf{x}')\,,\end{eqnarray}
obtained from the (\ref{sys11}) by setting $b_1=0$. In this case,
the Green function (Fourier transformed), is:

\begin{equation}\label{greenfunctionfR}\tilde{\mathcal{G}}_I(\mathbf{k})
=-\frac{2a_2Y_I^{-1}}{6a_2\mathbf{k}^4+a_1\mathbf{k}^2}\,,\end{equation}
and the Lagrangian becomes: $L=a_1R+a_2R^2$. Since at the level of
the Newtonian limit, as discussed, the powers of Ricci scalar
higher then two do not contribute, we can conclude that
(\ref{greenfunctionfR}) is the Green function for any $f(R)$ -
theory at Newtonian order, if $f(R)$ is some analytical function
of the Ricci scalar. The same result is achieved   considering a
particular choice of the  constants in the theory, e.g.
$b_1=-2a_2$. In Table \ref{tablefieldequation} (case iv), we
provide the field equations for this choice and the related Green
function is:

\begin{equation}\label{greenfunctionD}\tilde{\mathcal{G}}_{(2a_2\nabla^4-a_1\nabla^2)}(\mathbf{k})\propto\frac{1}{2a_2\mathbf{k}^4+a_1\mathbf{k}^2}\,.
\end{equation}
The spatial behavior of (\ref{greenfunctionfR}) -
(\ref{greenfunctionD}) is the same but the coefficients are
different since the theories are different. The interpretation of
the result is the same than that in $\S$
\ref{particularfieldequation} since we have to take into account a
proper scale length. In fact Eq.(\ref{greenfunctionfR}) presents a
null pole for $\textbf{k}^2\,=\,0$ which gives the standard
Newtonian potential, and a pole in $\textbf{k}^2\,=\,\lambda_2^2$,
which gives the Yukawa - like correction. Finally, we need the
Green function for the differential operator $\nabla^4$. From
Table \ref{tablefieldequation}, the field equations present always
a quadratic Laplacian operator (Case i excluded). This means that
the equation to solve is:

\begin{eqnarray}\label{laplaciaquadroequation}\nabla^4_{\mathbf{x}}
\mathcal{G}_{(\nabla^4)}(\textbf{x}',\textbf{x})\,=\delta(\textbf{x}-\textbf{x}')\,.
\end{eqnarray}
By introducing the variable $r\,=\,|\textbf{x}-\textbf{x}'|\,\neq
0$, we have that Eq.(\ref{laplaciaquadroequation}) becomes

\begin{eqnarray}\nabla^4_r\mathcal{G}_{(\nabla^4)}(r)\,=\,0\end{eqnarray}
with solution

\begin{eqnarray}\mathcal{G}_{(\nabla^4)}(r)\,=\,K_{I,1}+\frac{K_{I,2}}{r}+K_{I,3}r+K_{I,7}r^2\,,\end{eqnarray}
where $K_{I,1}$, $K_{I,2}$, $K_{I,3}$, $K_{I,7}$ are generic
integration constants.

Let us now consider the fact that the Green function has to be
null at infinity. The only possible physical choice for the
squared Laplacian is:

\begin{equation}\label{greenfunctionlaplsqu}\tilde{\mathcal{G}}_{(\nabla^4)}(\mathbf{x},\mathbf{x}')\propto\frac{1}{|\mathbf{x}-\mathbf{x}'|}\,.
\end{equation}
Considering the last possibility, we will end up with a force law
increasing with distance \cite{Havas:1977}. In conclusion, we have
shown the general approach to find solutions of the field
equations by using the Green functions. In particular, the vacuum
solutions with point-like source have been used to find out
directly the potentials, however it remains the most important
issue to find out  solutions  when we consider systems with
extended matter distribution.

\section{Solutions by  the Green functions in spherically symmetric distribution of matter}\label{sec:solgreen}

In this section, we explicitly determine the gravitational
potential in the inner and in the outer region of a spherically
symmetric matter distribution. This is a delicate problem since
the Gauss theorem is not valid for the gravity theories which we
are considering. In fact, in the Newtonian limit of General
Relativity, the equation for the gravitational potential,
generated by a point-like source

\begin{equation}\nabla_\mathbf{x}^2\mathcal{G}_{New. mech.}(\mathbf{x},\mathbf{x}')=-4\pi\delta(\mathbf{x}-\mathbf{x}')\end{equation}
is not satisfied by the new Green functions developed above. If we
consider the flux of force lines  $\mathbf{F}_{New. mech.}$
defined as

\begin{equation}\mathbf{F}_{New. mech.}:=-\frac{GM(\mathbf{x}-\mathbf{x}')}{|\mathbf{x}-\mathbf{x}'|^3}=-GM\mathbf{\nabla}_{\mathbf{x}}\mathcal{G}_
{New. mech.}(\mathbf{x},\mathbf{x}')\,,\end{equation} we obtain,
as standard,  the Gauss theorem:

\begin{equation}\int_\Sigma d\Sigma\,\,\,\,\mathbf{F}_{New. mech.}\cdot\hat{n}\propto M\,,\end{equation}
where $\Sigma$ is a generic two-dimensional surface and $\hat{n}$
its surface normal. The flux of field $\mathbf{F}_{New. mech.}$ on
the surface $\Sigma$ is proportional to the matter content $M$,
inside to the surface independently of the particular shape of
surface (Gauss theorem, or Newton theorem for the gravitational
field \cite{binney}). On the other hand, if we consider the flux
defined by the new Green function, its value is not proportional
to the enclosed mass but depends on the particular choice of the
surface:

\begin{equation}\label{nogauss}\int_\Sigma d\Sigma\,\,\,\,\mathbf{F}_{New. mech.}\cdot\hat{n}\propto M_\Sigma\,.\end{equation}
Hence $M_\Sigma$ is a mass-function depending on the surface
$\Sigma$. Then we have to find the solution inside/outside the
matter distribution by evaluating the quantity

\begin{equation}\int d^3\mathbf{x}'\mathcal{G}_I(\mathbf{x},\mathbf{x}')\rho(\mathbf{x}')\,,\end{equation}
and by imposing the boundary condition on the separation surface.

\subsection{The general solution by the Green function
$\mathcal{G}_I(\mathbf{x},\mathbf{x}')$}\label{gen_sol_green_function}

By considering the expressions (\ref{sol5}) and (\ref{sol6}) with
the Green function (\ref{green_function}) and by assuming
$\Phi(\mathbf{x})=0$, we have

\begin{eqnarray}\label{sol15}U(\mathbf{x})&=&4\pi GY_I\frac{(8a_2+3b_1)\nabla_\mathbf{x}^2-a_1}{a_1(2a_2+b_1)}\int d^3\mathbf{x}'
\mathcal{G}_I(\mathbf{x},\mathbf{x}')\rho(\mathbf{x}')\,,\end{eqnarray}

%\nonumber\\&=&(\mu_1+\mu_2\nabla_\mathbf{x}^2)\,\,G\int
%d^3\mathbf{x}'\mathcal{G}^A_I(\mathbf{x}-\mathbf{x}')\rho(\mathbf{x}')

\begin{eqnarray}\label{sol16}V(\mathbf{x})&=&4\pi
GY_I\frac{(4a_2+b_1)\nabla_\mathbf{x}^2-a_1}{a_1(2a_2+b_1)}\int
d^3\mathbf{x}'\mathcal{G}_I(\mathbf{x},\mathbf{x}')\rho(\mathbf{x}')\,.\end{eqnarray}
%\nonumber\\&=&(\mu_1+\mu_3\nabla_\mathbf{x}^2)\,\,G\int
%d^3\mathbf{x}'\mathcal{G}^A_I(\mathbf{x}-\mathbf{x}')\rho(\mathbf{x}')
%where ${\displaystyle \mu_1:=-\frac{4\pi
%Y_I}{2a_2+b_1}=-\frac{12\pi
%Y_I}{a_1}\frac{\lambda_1^2\lambda_2^2}{\lambda_1^2-\lambda_2^2}}$,
%${\displaystyle \mu_2:=\frac{4\pi
%Y_I(8a_2+3b_1)}{a_1(2a_2+b_1)}=\frac{4\pi
%Y_I}{a_1}\frac{4\lambda_1^2-\lambda_2^2}{\lambda_1^2-\lambda_2^2}}$,
%${\displaystyle \mu_3:=\frac{4\pi
%Y_I(4a_2+b_1)}{a_1(2a_2+b_1)}=\frac{4\pi
%Y_I}{a_1}\frac{2\lambda_1^2+\lambda_2^2}{\lambda_1^2-\lambda_2^2}}$.
We have to note that the hypothesis, $\Phi(\mathbf{x})=0$, is not
particular, since when we considered the Hilbert-Einstein
Lagrangian to give the Newtonian solution, we  imposed an
analogous condition. In fact, considering the spatial components
of the Einstein equations and then, by asking for $\Phi=0$, we get
the condition that the two metric potential $U$ and $V$ have to be
equal  (see Case i in Table \ref{tablefieldequation}). In the
general case, the role of $\Phi$ is given by $V-U$. From the time
- time component, we can obtain an expression for $U$. The next
step, in principle, is to search for a non-trivial solution
$\Phi$. This task is very difficult in general but can be realized
for some particular cases. For example,  since the Green function
can be found under the spherical symmetry hypothesis, also the
spatial distribution of matter have to be spherically symmetric.
Denoting the radius of the sphere with total mass $M$ by $\xi$, we
have the matter - density function

\begin{eqnarray}\rho(\mathbf{x})\,=\,\frac{3M}{4\pi\xi^3}\Theta(\xi-|\mathbf{x}|),\end{eqnarray}
where $\Theta(\xi-|\mathbf{x}|)$ is the Heaviside function. For
the potential $U(\mathbf{x})$, we obtain the implicit expression

\begin{eqnarray}U(\mathbf{x})&=&\frac{3GMY_I}{\xi^3}\frac{(8a_2+3b_1)\nabla_\mathbf{x}^2-a_1}{a_1(2a_2+b_1)}\times\nonumber\\&&\times
\int_0^\xi d|\mathbf{x}'||\mathbf{x}'|^2\int_0^{2\pi}
d\phi'\int_0^\pi
d\theta'\sin\theta'\mathcal{G}_I(\mathbf{x},\mathbf{x}')\,,\end{eqnarray}
and an analogous relation for $V(\mathbf{x})$. After some algebra,
we get the explicit form
\,%\,\,\,\,\,\,\,|\mathbf{x}|>\xi\,,\end{array}\right.\end{equation}
%gives us the internal and the external Newtonian behavior. The
%internal and the external potential for a given $\lambda_i$ is
%\begin{equation}U_{I,\,\,in}^{A,\lambda_i}(\mathbf{x})=\frac{3GM}{\lambda_i^2\xi^3}\biggl[\mu_1-e^{-\lambda_i\xi}(1+\lambda_i\xi)(\mu_1+
%\lambda_i^2\mu_2)\frac{\sinh(\lambda_i|\mathbf{x}|)}{\lambda_i|\mathbf{x}|}\biggr]\,,\end{equation}
%\begin{equation}U_{I,\,\,out}^{A,\lambda_i}(\mathbf{x})=\frac{3GM}{\lambda_i^2\xi^3}[\lambda_i\xi\cosh(\lambda_i\xi)-\sinh(\lambda_i\xi)](\mu_1+
%{\lambda_i}^2\mu_2)\frac{e^{-\lambda_i|\mathbf{x}|}}{\lambda_i|\mathbf{x}|}\,.\end{equation}

\begin{eqnarray}\label{potfinAin}U_{in}(\mathbf{x})&=&\frac{(2\pi)^{3/2}}{2a_1}\frac{GM}{\xi^3}\biggl[\frac{\lambda_1^2(2+3\lambda_2^2
\xi^2)-8\lambda_2^2}{\lambda_1^2\lambda_2^2}-|\mathbf{x}|^2\nonumber\\&&+8e^{-\lambda_1\xi}(1+\lambda_1\xi)\frac{\sinh(\lambda_1|\mathbf{x}|)}
{\lambda_1^3|\mathbf{x}|}\nonumber\\&&-2e^{-\lambda_2\xi}(1+\lambda_2\xi)\frac{\sinh(\lambda_2|\mathbf{x}|)}{\lambda_2^3|\mathbf{x}|}\biggr]
\,,\end{eqnarray}

\begin{eqnarray}\label{potfinAout}U_{out}(\mathbf{x})&=&\frac{(2\pi)^{3/2}}{a_1}\frac{GM}{|\mathbf{x}|}\nonumber\\&&-\frac{4(2\pi)^{3/2}}{a_1}\frac{GM}
{\lambda_1^3\xi^3}[\lambda_1\xi\cosh(\lambda_1\xi)-\sinh(\lambda_1\xi)]\frac{e^{-\lambda_1|\mathbf{x}|}}{|\mathbf{x}|}\nonumber\\&&
+\frac{(2\pi)^{3/2}}{a_1}\frac{GM}{\lambda_2^3\xi^3}[\lambda_2\xi\cosh(\lambda_2\xi)-\sinh(\lambda_2\xi)]\frac{e^{-
\lambda_2|\mathbf{x}|}}{|\mathbf{x}|}.\end{eqnarray} The relations
(\ref{potfinAin})-(\ref{potfinAout}) give the solutions for the
gravitational potential $U$ inside and outside the constant
spherically symmetric matter distribution. A similar relation is
found for $V(\textbf{x})$. The boundary condition on the surface
$|\mathbf{x}|=\xi$ is satisfied:

\begin{equation}\label{bound_cond}U_{in}(\xi)-U_{out}(\xi)\,=\,0\,.\end{equation}
We note that the corrections to the Newtonian terms are ruled by
$\mathcal{G}_I(\mathbf{x},\mathbf{x}')$. In order to achieve the
behavior of the potential inside the matter distribution, let us
perform a Taylor expansion for $\lambda_i|\mathbf{x}|\ll 1$. We
have:

\begin{equation}\label{senoiper}\frac{\sinh(\lambda|\mathbf{x}|)}{\lambda|\mathbf{x}|}\simeq constant+|\mathbf{x}|^2+...\,
.\end{equation} Like for the standard Newtonian potential, this
means that the inner solution of the corrected potential is traced
by the matter distribution.

The outer solution has to be discussed in detail. For fixed values
of the distance $|\mathbf{x}|$, the external potential
$U_{out}(\mathbf{x})$ depends on the value of the radius $\xi$,
then the Gauss theorem does not work also if the Bianchi
identities hold \cite{Stelle:1978}. In other words, since the
Green function does not scale as the inverse distance but has an
exponential behavior, the  Gauss theorem (\ref{nogauss}) does not
hold. This means that \emph{the potential depends on the total
mass and on the matter - distribution in the space}. In
particular, if the matter distribution takes a bigger volume, the
potential $|U_{out}(\mathbf{x})|$ increases and viceversa. We can
write

\begin{equation}\label{geometric_fattor}\lim_{\xi\rightarrow\infty}\frac{\lambda\xi\cosh(\lambda\xi)-\sinh(\lambda\xi)}{\lambda^3\xi^3}=\infty\,;
\end{equation}
obviously the limit of $\xi$ has to be considered up to the
maximal value of $|\mathbf{x}|$. The term defined in
(\ref{geometric_fattor}) can be defined as a sort of geometric
factor which takes into account the spatial matter-distribution.
The limit puts in evidence the dependence on the matter
distribution of the outer potential. The spherical symmetry allows
to find out  the Green functions but, in principle, they can be
achieved also without this hypothesis.  At this point, it is
interesting to consider the physical meaning of the Green
functions. Eq.(\ref{green_function}) represents the correlation
(i.e. the interaction) between two points in the space, that is,
it gives the possibility to calculate the potential in a given
point as a function of the "charge" (the point - mass) of another
point. Such a charge is described by a Dirac delta function which
is the source term of the Green function. By summing up the
contributions of all infinitesimal volume elements, we obtain the
potential for a given matter - distribution. This analysis can be
concluded with some considerations related to the behavior for
$|\textbf{x}|\gg\xi$. This means that we are moving away from the
matter distribution. Such a limit can be given also as
$\xi\rightarrow 0$, that is:

\begin{equation}\lim_{\xi\rightarrow
0}3\frac{\lambda\xi\cosh(\lambda\xi)-\sinh(\lambda\xi)}{\lambda^3\xi^3}=1\,.
\end{equation}
For $U_{out}(\mathbf{x})$, we have

\begin{eqnarray}\lim_{\xi\rightarrow 0}\,\,U_{out}(\mathbf{x})&=&\frac{(2\pi)^{3/2}}{a_1}\frac{GM}{|\mathbf{x}|}-\frac{4(2\pi)^{3/2}}{3a_1}
\frac{GMe^{-\lambda_1|\mathbf{x}|}}{|\mathbf{x}|}\nonumber\\&&+\frac{(2\pi)^{3/2}}{3a_1}\frac{GMe^{-\lambda_2|\mathbf{x}|}}{|\mathbf
{x}|}\,.\end{eqnarray} We can choose $a_1\,=\,(2\pi)^{3/2}$, and
then, for a point - like mass, we have:

\begin{eqnarray}\label{outpointlike1}\lim_{\xi\rightarrow 0}\,\,U_{out}(\textbf{x})\,=\,\frac{GM}{|\mathbf{x}|}-\frac{4}{3}\frac{GMe^{-\lambda_1
|\mathbf{x}|}}{|\mathbf{x}|}+\frac{1}{3}\frac{GMe^{-\lambda_2|\mathbf{x}|}}{|\mathbf{x}|}\,.\end{eqnarray}
The last expression is compatible with the discussion in $\S$.
\ref{sec:greenfunctparagraph}.

\subsection{Further solutions by the Green functions $\mathcal{G}_I(\mathbf{x},\mathbf{x}')$}

For the sake of completeness, let us derive  the expression for
the potential $U(\textbf{x})$ for the other two Green functions in
Table \ref{tablegrennfunction}. By performing a similar
calculation, but now  considering Case ix in Table
\ref{tablegrennfunction}, we obtain

%\begin{equation}\mathcal{H}^{B,\,\lambda_i}_I(\mathbf{x})=\left\{\begin{array}{ll}\frac{3GM}{\lambda_i^2\xi^3}\biggl\{-1+[\cos(\lambda_i\xi)+\lambda_i
%\xi\sin(\lambda_i\xi)]
%\frac{\sin(\lambda_i|\mathbf{x}|)}{\lambda_i|\mathbf{x}|}\biggr\}\,\,\,|\mathbf{x}|<\xi\\\\\frac{3GM}{{\lambda_i}^2\xi^3}[\sin(\lambda_i\xi)-\lambda_
%i\xi\cos(\lambda_i\xi)]\frac{\cos(\lambda_i|\mathbf{x}|)}{\lambda_i|\mathbf{x}|}\,\,\,\,\,\,\,\,\,\,\,\,\,\,\,\,\,\,\,\,\,\,\,\,|\mathbf{x}|>\xi\,,
%\end{array}\right.
%\end{equation} in the inner and outer region. Also in this case, if
%we consider the limit of $\lambda_i\rightarrow 0$, one obtains the
%Newtonian limit (\ref{newlimit}). The internal and external
%potential for given $\lambda_i$ is
%\begin{eqnarray}U_{I,\,\,in}^{B,\lambda^i}(\mathbf{x})&=&\frac{3GM}{\lambda_i^2\xi^3}\biggl\{-\mu_1+(\mu_1-{\lambda_i}^2\mu_2)[\cos(\lambda_i\xi)+
%\nonumber\\&&
%+\lambda_i\xi\sin(\lambda_i\xi)]\frac{\sin\lambda_i|\mathbf{x}|}{\lambda_i|\mathbf{x}|}\biggr\}\,,\end{eqnarray}
%\begin{equation}U_{I,\,\,out}^{B,\lambda_i}(\mathbf{x})=\frac{3GM}{\lambda_i^2\xi^3}(\mu_1-{\lambda_i}^2\mu_2)[\sin(\lambda_i\xi)-\lambda_i\xi\cos(
%\lambda_i\xi)]\frac{\cos(\lambda_i|\mathbf{x}|)}{\lambda_i|\mathbf{x}|}\,.\end{equation}
%The boundary condition on the surface $|\mathbf{x}|=\xi$ is
%\begin{equation}\label{boundcodB}_BU_{I,\,\,in}(\xi)-_BU_{I,\,\,out}(\xi)=-\frac{3GM}{\xi^3}\mu_2\sum_{i=0}^3\mathcal{G}^B_i=0\,,\end{equation}
%$(see Table \ref{tablegrennfunction}). The internal and external potential are given by

\begin{eqnarray}\label{potfinBin}U_{in}(\mathbf{x})&=&\frac{GM}{2\xi^3}\biggl\{\frac{\lambda_1^2(3\lambda_2^2\xi^2
-2)+8\lambda_2^2}{\lambda_1^2\lambda_2^2}-|\mathbf{x}|^2\nonumber\\&&-\frac{8}{\lambda_1^2}[\cos(\lambda_1\xi)+\lambda_1\xi\sin(\lambda_1\xi)]\frac{\sin(
\lambda_1|\mathbf{x}|)}{\lambda_1|\mathbf{x}|}\nonumber\\&&+\frac{2}{\lambda_2^2}[\cos(\lambda_2\xi)+\lambda_2\xi\sin(\lambda_2\xi)]\frac{\sin(\lambda_2
|\mathbf{x}|)}{\lambda_2|\mathbf{x}|}\biggr\}\,,\end{eqnarray}

\begin{eqnarray}\label{potfinBout}U_{out}(\mathbf{x})&=&\frac{GM}{|\mathbf{x}|}-\frac{4(2\pi)^{3/2}}{a_1}
\frac{GM}{\lambda_1^3\xi^3}[\sin(\lambda_1\xi)\nonumber\\&&-\lambda_1\xi\cos(\lambda_1\xi)]\frac{\cos(\lambda_1|\mathbf{x}|)}{|\mathbf{x}|}+\frac{(2\pi)
^{3/2}}{a_1}\frac{GM}{\lambda_2^3\xi^3}[\sin(\lambda_2\xi)\nonumber\\&&-\lambda_2\xi\cos(\lambda_2\xi)]\frac{\cos(\lambda_2|\mathbf{x}|)}{|\mathbf{x}|}
\biggr]\,.\end{eqnarray} Also in this case the boundary conditions
(\ref{bound_cond}) are satisfied. The  considerations of preceding
subsection hold also for the solutions (\ref{potfinBin}) -
(\ref{potfinBout}). The only difference is that now we have
oscillating behaviors instead of exponential behaviors. The
correction term to the Newtonian potential in the external
solution  can be interpreted as the Fourier transform of the
matter density $\rho(\mathbf{x})$. In fact, we have:

\begin{equation}\int\frac{d^3\mathbf{x}'}{(2\pi)^{3/2}}\rho(\mathbf{x}')e^{-i\mathbf{k}\cdot\mathbf{x}'}=\frac{3M}{(2\pi)^{2/3}}
\frac{\sin(|\mathbf{k}|\xi)-|\mathbf{k}|\xi\cos(|\mathbf{k}|\xi)}{|\mathbf{k}|^3\xi^3}\,,\end{equation}
and in the point - like mass limit, it is

\begin{equation}\lim_{\xi\rightarrow 0}\int\frac{d^3\mathbf{x}'}{(2\pi)^{3/2}}\rho(\mathbf{x}')e^{-i\mathbf{k}\cdot\mathbf{x}'}=\frac{M}{(2\pi)^{2/3}}\,,
\end{equation}
we obtain again the external solution for point-like source as
limit of (\ref{potfinBout}):

\begin{eqnarray}\label{outpointlike2}\lim_{\xi\rightarrow 0}\,\,U_{out}(\mathbf{x})&=&\frac{GM}{|\mathbf{x}|}-\frac{2}{3}\frac{GM\cos(\lambda_1
|\mathbf{x}|)}{|\mathbf{x}|}+\frac{1}{6}\frac{GM\cos(\lambda_2|\mathbf{x}|)}{|\mathbf{x}|}
\,.\end{eqnarray}  Finally for the last case in Table
\ref{tablegrennfunction}, we have

\begin{eqnarray}U_{in}(\mathbf{x})&=&\frac{GM}{2\xi^3}\biggl\{\frac{\lambda_1^2(3\lambda_2^2\xi^2-2)-8\lambda_2}{
\lambda_1^2\lambda_2^2}-|\mathbf{x}|^2\nonumber\\&&+\frac{8}{\lambda_1^2}e^{-\lambda_1\xi}(1+\lambda_1\xi)\frac{\sinh(\lambda_1|\mathbf{x}|)}
{\lambda_1|\mathbf{x}|}\nonumber\\&&+\frac{2}{\lambda_2^2}[\cos(\lambda_2\xi)+\lambda_2\xi\sin(\lambda_2\xi)]\frac{\sin(\lambda_2|\mathbf{x}
|)}{\lambda_2|\mathbf{x}|}\biggl\}\,,\end{eqnarray}

\begin{eqnarray}U_{out}(\mathbf{x})&=&\frac{GM}{|\mathbf{x}|}-\frac{4(2\pi)^{3/2}}{a_1}\frac{GM}{
\lambda_1^3\xi^3}\nonumber\\&&\times[\lambda_1\xi\cosh(\lambda_1\xi)-\sinh(\lambda_1\xi)]\frac{e^{-\lambda_1|\mathbf{x}|}}{|\mathbf{x}|}\nonumber\\&&+
\frac{(2\pi)^{3/2}}{a_1}\frac{GM}{\lambda_2^3\xi^3}\nonumber\\&&\times[\sin(\lambda_2
\xi)-\lambda_2\xi\cos(\lambda_2\xi)]\frac{\cos(\lambda_2|\mathbf{x}|)}{|\mathbf{x}|}\,.\end{eqnarray}
The limit of point-like source is valid also in this  case, that
is:

\begin{eqnarray}\label{outpointlike3}\lim_{\xi\rightarrow 0}\,\,U_{out}(\mathbf{x})&=&\frac{GM}{|\mathbf{x}|}-\frac{4}{3}\frac{GMe^{-\lambda_1
|\mathbf{x}|}}{|\mathbf{x}|}+\frac{1}{3}\frac{GM\cos(\lambda_2|\mathbf{x}|)}{|\mathbf{x}|}\,.\end{eqnarray}
The results (\ref{outpointlike1}) - (\ref{outpointlike2}) -
(\ref{outpointlike3}) means that, for suitable distance scales,
the Gauss theorem is recovered and the theory agrees with the
standard Newtonian limit of General Relativity.

\subsection{Other solutions and their physical consistency}\label{particularsolution}

In Table \ref{tablefieldequationsolution}, we provide solutions,
in terms of the Green function of the corresponding differential
operator, for the field equations shown in Table
\ref{tablefieldequation}. Case i corresponds to the Newtonian
theory and the arbitrary constant $a_1$ can be absorbed in the
definition of matter Lagrangian as above. The implicit solution
is:

\begin{equation}U(\mathbf{x})=V(\mathbf{x})=G\int d^3\mathbf{x}'\frac{\rho(\mathbf{x}')}{|\mathbf{x}-\mathbf{x}'|}\,.\end{equation}
For  Case iv,  we have:

\begin{equation}\label{potsolD}U(\mathbf{x})=V(\mathbf{x})=G\int d^3\mathbf{x}'\biggl[\frac{1-e^{-\sqrt{\frac{a_1}{2a_2}}|\mathbf{x}-
\mathbf{x}'|}}{|\mathbf{x}-\mathbf{x}'|}\biggr]\rho(\mathbf{x}')\,.\end{equation}
The solutions make sense only if $a_1/a_2>0$, which gives a
scale-length. Also in this case, we can have different signatures
for  $a_1$ and $a_2$ which give  oscillating corrections to the
Newtonian potential.  We have to note that both above cases have
been solved with the hypothesis $\Phi\,=\,0$. These two cases are
the only ones which exhibit the standard Newtonian limit
(obviously the former). The remaining cases can exhibit
divergences and incompatibilities. This is obvious since, as
discussed in $\S$ \ref{particularfieldequation}, the absence in
the Lagrangians of terms linear in the Ricci curvature scalar
gives field equations with higher - order Laplacian operators (see
Cases ii, iii, v, vi, and vii). Precisely, without terms like
$\nabla^2 U + ... = \rho$, we could not achieve regular
Newtonian-like behaviors. This fact could give problems in
comparing inner and outer solutions with respect to  matter
distributions. In other words, the Newtonian potential is
necessary not only to achieve physically interesting situations
but also, from a mathematical point of view, to regularize
solutions. In fact, Case ii presents an incompatibility between
the solution obtained from the $00$ - component and the one from
the $ab$-component. The incompatibility can be removed if we
consider, as the Green function for the differential operator
$\nabla^4$, the trivial solution:
$\mathcal{G}_{(\nabla^4)}|_B=const.$ Only with this  choice, the
arbitrary integration constant $U_0$ can be interpreted as $GM$.
However another problem remains: namely the divergence in the
origin and then we can conclude that the solution
\begin{equation}2V(\mathbf{x})-U(\mathbf{x})=\frac{GM}{|\mathbf{x}-\mathbf{x}'|}\end{equation}
holds only in vacuum.

Besides,  terms like $\int
d^3\mathbf{x}'\mathcal{G}_{(\nabla^4)}(\mathbf{x},\mathbf{x}')\rho(\mathbf{x}')$
have to be discussed for the  choice (\ref{greenfunctionlaplsqu}).
The field equation with $\nabla^4$ (see Table
\ref{tablefieldequation}) gives

\begin{equation}\label{inconsi}\nabla_\mathbf{x}^4U(\mathbf{x})\propto\nabla_\mathbf{x}^4\int d^3\mathbf{x}'\frac{\rho(\mathbf{x}')}{|\mathbf{x}-
\mathbf{x}'|}=-4\pi\nabla^2_\mathbf{x}\rho(\mathbf{x})\neq-4\pi\rho(\mathbf{x});\end{equation}
which is  consistent only if $\rho(\mathbf{x})=0$. Due to these
considerations, also in the remaining cases, we can consistently
consider only vacuum solutions.

\begin{table}[ht]
 \centering
 \begin{tabular}{c|c|c}
 \hline\hline\hline
 Cases & Solutions & Newtonian behavior \\
 \hline
 & & \\
 i & $\begin{array}{ll}
 U(\mathbf{x})=V(\mathbf{x})=\frac{G}{a_1}\int
 d^3\mathbf{x}'\frac{\rho
 (\mathbf{x}')}{|\mathbf{x}-\mathbf{x}'|}\end{array}$ & yes \\
 \hline
 & & \\
 ii & $\begin{array}{ll}
 2V(\mathbf{x})-U(\mathbf{x})=\frac{U_0}{|\mathbf{x}|}\\\\ 2V(\mathbf{x})-U(\mathbf{x})=\frac{2\pi G}{a_2}\int
 d^3\mathbf{x}'\mathcal{G}_{(\nabla^4)}(\mathbf{x},\mathbf{x}')\rho
 (\mathbf{x}')\end{array}$ & no \\
 \hline
 & & \\
 iii & $\begin{array}{ll}
 U(\mathbf{x})=\frac{U_0}{|\mathbf{x}|}-\frac{6\pi G}{b_1}\int
 d^3\mathbf{x}'\mathcal{G}_{(\nabla^4)}(\mathbf{x},\mathbf{x}')\rho
 (\mathbf{x}')\\\\V(\mathbf{x})=\frac{U_0}{|\mathbf{x}|}-\frac{2\pi G}{b_1}\int
 d^3\mathbf{x}'\mathcal{G}_{(\nabla^4)}(\mathbf{x},\mathbf{x}')\rho
 (\mathbf{x}')\end{array}$ & no \\
 \hline
 & & \\
 iv & $\begin{array}{ll}
 U(\mathbf{x})=4\pi G\int d^3\mathbf{x}'\mathcal{G}_{(2a_2\nabla^4-a_1\nabla^2)}(\mathbf{x},\mathbf{x}')\rho
 (\mathbf{x}') \\\\V(\mathbf{x})=4\pi G\int d^3\mathbf{x}'\mathcal{G}_{(2a_2\nabla^4-a_1\nabla^2)}(\mathbf{x},\mathbf{x}')\rho
 (\mathbf{x}')\end{array}$ & yes \\
 \hline
 & & \\
 v & $\begin{array}{ll}
 U(\mathbf{x})=\frac{2\pi G}{a_2}\int
 d^3\mathbf{x}'\mathcal{G}_{(\nabla^4)}(\mathbf{x},\mathbf{x}')\rho
 (\mathbf{x}')\\\\V(\mathbf{x})=\frac{U_0}{|\mathbf{x}|}\end{array}$ & no\\
 \hline
 & & \\
 vi & $\begin{array}{ll}
 U(\mathbf{x})=\frac{U_0}{|\mathbf{x}|}+\frac{2\pi G}{a_2}\int
 d^3\mathbf{x}'\mathcal{G}_{(\nabla^4)}(\mathbf{x},\mathbf{x}')\rho
 (\mathbf{x}')\\\\V(\mathbf{x})=\frac{2\pi G}{a_2}\int
 d^3\mathbf{x}'\mathcal{G}_{(\nabla^4)}(\mathbf{x},\mathbf{x}')\rho
 (\mathbf{x}')\end{array}$ & no\\
 \hline
 & & \\
 vii & $\begin{array}{ll}
 U(\mathbf{x})=\frac{U_0}{|\mathbf{x}|}
 \\\\V(\mathbf{x})=-\frac{1}{2}\frac{U_0}{|\mathbf{x}|}+\frac{3\pi G}{a_2}\int
 d^3\mathbf{x}'\mathcal{G}_{(\nabla^4)}(\mathbf{x},\mathbf{x}')\rho
 (\mathbf{x}')\end{array}$ & no\\
 \hline\hline\hline
 \end{tabular}
 \caption{\label{tablefieldequationsolution}Here we provide the
 solutions of the field equations in Table
 \ref{tablefieldequation}. The solutions are found by setting
 $\Phi(\mathbf{x})=0$ in the $ab$\,-\,component of the field equation (\ref{sys12})
 or (\ref{sys22}). The solutions are displayed in terms
 of the Green functions. $U_0$ is a generic integration constant.}
\end{table}

\section{Conclusions and outlook}
\label{sec:conclusions:outlook}

In this paper, we have studied the Newtonian limit of
gravitational theories whose action presents quadratic curvature
invariants beside the standard Ricci curvature scalar of General
Relativity. In particular, we have considered the problem to find
out solutions of the field equations developed up to the
perturbation order  ${\mathcal O}\left(c^{-2} \right)$. This is
intended as the Newtonian limit while, taking into account terms
up to ${\mathcal O}\left(c^{-3} \right)$ and beyond is the
post-Newtonian approximation (see for example \cite{gravmagn}).

After  deriving the full fourth - order field equations, we have
developed the metric and the stress - energy tensors in the
Newtonian limit.  The main metric quantities, in this limit, are
the two gravitational potentials $U$ and $V$ which are the
solutions of the field equations both in presence and in absence
of matter. At the order $c^{-2}$, quadratic curvature invariants
give rise to $\nabla^2$ and $\nabla^4$ operators acting on $U$ and
$V$ in the field equations. Our task has been to develop an
approach to solve such equations and to find out corrections to
the  Newtonian potential emerging, as standard, from General
Relativity.

The method consists in searching for suitable combinations of the
gravitational potentials $U$ and $V$ by which it is possible to
decouple the field equations. After field equations are suitably
decoupled, one can define Green's functions which allow to obtain
the potentials. These potentials, however, strictly depend on the
coupling parameters appearing in the Lagrangian of the  theory.
Conversely, such coupling constants allow to classify the field
equations, and then the solutions, selecting, in particular, some
singular cases.

A detailed discussion has been developed for systems presenting
spherical symmetry. In this case, the role of corrections to the
Newtonian potential is clearly evident. In general, such
corrections are oscillating behaviors or Yukawa-like terms. This
means that one of the effects to introduce quadratic curvature
invariants is to select characteristic scale lengths which could
have physical interests as we will discuss below. Besides, such
corrections invalidate the Gauss theorem because any matter
distribution depends on such scale lengths. Furthermore, if the
Newtonian potential term is not present, there could be
compatibility problems and some solutions are physically
consistent only in vacuum. Furthermore, for spherically symmetric
distributions of matter, we discussed the inner and the outer
solutions and the boundary conditions.

From a physical viewpoint, this systematic work is needed in order
to fully develop the weak field limit of such relativistic
theories of gravity and then compare them with observations and
experiments. In fact, the correct interpretation of data  strictly
depends on the self-consistency of the theory and, viceversa, data
correctly interpreted could definitively confirm or rule out
deviations from General Relativity \cite{will}. It is worth
pointing out that extended or alternative theories of gravity seem
good candidates to solve several shortcomings of modern
astrophysics and cosmology since they could address several issues
of cosmological dynamics without introducing unknown forms of dark
matter and dark energy (see e.g.\cite{JCAP,Capozziello:2002}).
Nevertheless, a "final" alternative theory solving all the issues
has not been found out up to now and the debate on modifying
gravitational sector or adding new (dark) ingredients is still
open. Beside this general remark related to the paradigm
(extending gravity and/or adding new components), there is the
methodological issue to "recover" the standard and well-tested
results of General Relativity in the framework of these
alternative schemes. The recovering of a self-consistent Newtonian
limit is the test bed of any theory of gravity which pretends to
enlarge or correct the Einstein General Relativity.

Taking  into account also the results presented in
\cite{Newton:2007,Curv:2007}, it is clear that only General
Relativity presents directly the Newtonian potential in the weak
field limit while corrections (e.g. Yukawa-like terms) appear as
soon as the theory is non-linear in the Ricci scalar. This
occurrence could be particularly useful to solve the problem of
missing matter in large astrophysical systems like galaxies and
clusters of galaxies as discussed in \cite{MNRAS,MNRAS1}. In fact
dark matter (and dark energy) could be nothing else but the
effects that General Relativity, experimentally tested only up to
Solar System scales, does not work at extragalactic scales and
then it has to be corrected. Assuming this alternative point of
view, we do not need to search for unknown ingredients, up to now
not found at fundamental level, but we need only to revise the
behavior of gravitational field at infrared scales. These scales
could be ruled by corrections to the Newtonian potential, as shown
in this paper. In forthcoming researches, we intend to confront
such solutions with experimental data, as done in
\cite{MNRAS,MNRAS1}, in order to see if large self-gravitating
systems could be modelled by them.

\section*{Acknowledgements}
The Authors warmly thank A. Troisi for useful discussions and
comments on the topic.  

%\clearpage

\newpage

\appendix

\section{Alternative approach to solve the field
equations}\label{app:approach}

In this Appendix, we discuss an alternative approach to solve the
field equations where, instead of using the relation
$A=\Phi-\nabla^2B$ as in $\S$ \ref{general_approach} to obtain the
solutions (\ref{sol5}) - (\ref{sol6}), we adopt the inverse
relation $\nabla^2B=\Phi-A$. As noted above, the first relation
makes the differential degree of system increase but we have a
relation between the solutions $A$ and $B$. In the second case the
differential degree remains the same but we have a non-local
relation between the solutions:
\begin{equation}B(\mathbf{x})=\frac{1}{4\pi}\int
d^3\mathbf{x}'\frac{A(\mathbf{x}')-\Phi(\mathbf{x}')}{|\mathbf{x}-\mathbf{x}'|}\,.
\end{equation}
In this  case, the boundary conditions play a crucial role in the
integration process. In general, considering the inverse relation
$\nabla^2B=\Phi-A$, we have a new set of equations
\begin{equation}\label{eq7}\frac{2b_1(3a_2+b_1)}{a_1(2a_2+b_1)}\nabla^4A+\frac{6a_2+b_1}{2a_2+b_1}\nabla^2A-\frac{a_1}{2a_2
+b_1}A =8\pi G\rho-\tau_{II}\,,\end{equation}
\begin{equation}\label{eq8}\nabla^2\Phi\delta_{ab}-\Phi_{,ab}=0\,,\end{equation} where
${\displaystyle \tau_{II}:=-\frac{a_1}{2a_2+b_1}\Phi+\frac{2a_2}
{2a_2+b_1}\nabla^2\Phi}$. Now by introducing a new Green function
$\mathcal{G}_{II}(\mathbf{x},\mathbf{x}')$, we have
\begin{equation}\label{gr2}A(\mathbf{x})=Y_{II}\int d^3\mathbf{x}'\mathcal{G}_{II}
(\mathbf{x},\mathbf{x}')\sigma_{II}(\mathbf{x}')\,,\end{equation}
where
\begin{equation}\label{den2}\sigma_{II}(\mathbf{x}):=8\pi G\rho(\mathbf{x})-\tau_{II}(\mathbf{x})\,,\end{equation} and
$Y_{II}$ is a constant having the dimension of a length$^{-1}$.
Eqs.(\ref{eq:combined:1}) - (\ref{eq:combined:3}) become
\begin{eqnarray}\label{sys21}&
\frac{2b_1(3a_2+b_1)}{a_1(2a_2+b_1)}\nabla^4_{\mathbf{x}}\mathcal{G}_{II}(\mathbf{x},\mathbf{x}')+\frac{6a_2+b_1}{2a_2+b_1}\nabla^2_{\mathbf{x}}
\mathcal{G}_{II}(\mathbf{x},\mathbf{x}')\nonumber\\&-\frac{a_1}{2a_2+b_1}\mathcal{G}_{II}(\mathbf{x},\mathbf{x}')=Y_2^{-1}\delta(\mathbf{x}-\mathbf{x}')
\,,\end{eqnarray}
\begin{equation}\label{sys22}\nabla^2\Phi(\mathbf{x})\delta_{ab}-\Phi(\mathbf{x})_{,ab}=0\,.\end{equation}
The general solutions of Eqs.(\ref{eq:combined:1}) -
(\ref{eq:combined:3}), by introducing the Green function
$\mathcal{G}_{II}(\mathbf{x},\mathbf{x}')$ and the function
$\Phi(\mathbf{x})$, are
\begin{eqnarray}\label{sol7}U(\mathbf{x})&=&-\frac{8a_2+3b_1}{2a_1(2a_2+b_1)}Y_{II}\int d^3\mathbf{x}'\mathcal{G}_{II}(\mathbf{x},\mathbf{x}')\biggl[
8\pi
G\rho(\mathbf{x}')\nonumber\\&&+\frac{a_1}{2a_2+b_1}\Phi(\mathbf{x}')-\frac{2a_2}{2a_2+b_1}\nabla_{\mathbf{x}'}^2\Phi(\mathbf{x}')\biggr]\nonumber\\&&+
\frac{Y_{II}}{8\pi(2a_2+b_1)}\int
d^3\mathbf{x}'d^3\mathbf{x}''\frac{\mathcal{G}_{II}(\mathbf{x}',\mathbf{x}'')}{|\mathbf{x}-\mathbf{x}'|}\biggl[8\pi
G\rho(\mathbf{x}'')\nonumber\\&&+\frac{a_1}{2a_2+b_1}\Phi(\mathbf{x}'')-\frac{2a_2}{2a_2+b_1}\nabla_{\mathbf{x}''}^2\Phi(\mathbf{x}'')\biggr]\nonumber\\
&&-\frac{1}{8\pi(2a_2+b_1)}\int
d^3\mathbf{x}'\frac{\Phi(\mathbf{x}')}{|\mathbf{x}-\mathbf{x}'|}\,,\end{eqnarray}
\begin{eqnarray}\label{sol8}V(\mathbf{x})&=&-\frac{(4a_2+b_1)}{2a_1(2a_2+b_1)}Y_{II}\int d^3\mathbf{x}'\mathcal{G}_{II}(\mathbf{x},\mathbf{x}')\biggl[
8\pi
G\rho(\mathbf{x}')\nonumber\\&&+\frac{a_1}{2a_2+b_1}\Phi(\mathbf{x}')-\frac{2a_2}{2a_2+b_1}\nabla_{\mathbf{x}'}^2\Phi(\mathbf{x}')\biggr]\nonumber\\&&-
\frac{Y_{II}}{4\pi a_1(2a_2+b_1)}\int
d^3\mathbf{x}'d^3\mathbf{x}''\frac{\mathcal{G}_{II}(\mathbf{x}',\mathbf{x}'')}{|\mathbf{x}-\mathbf{x}'|}\biggl[8\pi
G\rho(\mathbf{x}'')\nonumber\\&&+\frac{a_1}{2a_2+b_1}\Phi(\mathbf{x}'')-\frac{2a_2}{2a_2+b_1}\nabla_{\mathbf{x}''}^2\Phi(\mathbf{x}'')\biggr]\nonumber\\
&&+\frac{1}{4\pi a_1(2a_2+b_1)}\int
d^3\mathbf{x}'\frac{\Phi(\mathbf{x}')}{|\mathbf{x}-\mathbf{x}'|}\,.\end{eqnarray}
With the second relation between $A$ and $B$, Eq.(\ref{eq7}) is a
fourth - order equation, but in this case, the potentials $U$, $V$
are linked to $A$ through repeated integrations of (\ref{sol7}) -
(\ref{sol8});  for the first choice, we have only the integral
(\ref{sol5}) - (\ref{sol6}). If we consider, instead,
Eq.(\ref{sys21}), we can find a similar Green function
$\mathcal{G}_{II}(\mathbf{x},\mathbf{x}')$ for the solutions
(\ref{sol7})-(\ref{sol8}). In fact, for $r\neq 0$, we have
\begin{equation}\label{eq14}2b_1(3a_2+b_1)\nabla_r^4\mathcal{G}_{II}(r)+a_1(6a_2+b_1)\nabla_r^2\mathcal{G}_{II}(r)-a^2_1
\mathcal{G}_{II}(r)=0\,,\end{equation} and its solution is similar
to (\ref{sol9}):
\begin{eqnarray}\label{sol4}\mathcal{G}_{II}(r)&=&\frac{1}{r}\biggl[K_{II,3}e^{-\sqrt{-\frac{a_1}{b_1}}r}
+K_{II,4}e^{\sqrt{-\frac{a_1}{b_1}}r}+K_{II,5}e^{-\sqrt{\frac{a_1}{2(3a_2+b_1)}}r}\nonumber\\&&+
K_{II,6}e^{\sqrt{\frac{a_1}{2(3a_2+b_1)}}r}\biggr]\,,\end{eqnarray}
where, as above, $K_{II,3}$, $K_{II,4}$, $K_{II,5}$, $K_{II,6}$
are  constants. It is worth noticing that  it is not possible to
factorize the Laplacian in (\ref{sys21}) and, in terms of Fourier
transform, a vanishing pole is not present. If the
Fourier transformed function has no pole in the origin ($k^2 =
0$), this means that, in the field equation, we do not have all
terms containing a Laplacian but, some term can be interpreted as
the mass for the field.  On the other hand, if it is not possible
to factorize a Laplacian in the field equation, this means that
the pole $k^2 = 0$ is absent in the Fourier transform of the Green
function and a Newtonian potential scaling as $r^-1$ is not
present.
%It is
%\begin{equation}\label{intfou2}\mathcal{G}_{II}(\mathbf{x}-\mathbf{x}')=Y^{-1}_{II}\int\frac{d^3\mathbf{k}}{(2\pi)^{3/2}}\frac{e^{i\mathbf{k}\cdot(
%\mathbf{x}-\mathbf{x}')}}{\alpha_1\mathbf{k}^4-\alpha_2\mathbf{k}^2+\alpha_3}\,\end{equation}
Let us remember that a potential scaling like $1/r$ is used in the
Green function proportional to $\int
d^3\mathbf{k}\,\mathbf{k}^{-2}\,e^{i\mathbf{k}\cdot(\mathbf{x}-
\mathbf{x}')}$. Furthermore, the analogy between the two
approaches is complete when we consider the link between them
being:
\begin{equation}\mathcal{G}_{II}(\mathbf{x},\mathbf{x}')=\nabla^2_{\mathbf{x}}\mathcal{G}_{I}(\mathbf{x},\mathbf{x}')\,.\end{equation}
%Obviously, if we consider the Green function
%$\mathcal{G}_{II}(\mathbf{x}-\mathbf{x}')$, the potentials $U$ and
%$V$ will be expressed by (\ref{sol7})-(\ref{sol8}).

\section{Conventions $\&$ Dimensions}\label{app:conventions}

In order to fix the notation, we provide two tables with
definitions. Table \ref{tab:definitions} gives an overview of the
geometrical quantities used in  the paper. We make use of the
summation convention over identical upper and lower indices
throughout the paper. The dimensions of the different quantities
appearing throughout the work are displayed in Table
\ref{tab:dimensions}.

%\newpage

\begin{table}

\begin{tabular}{ll}
\hline\hline\hline
Object&Definition/Convention\\
\hline & \\
Index ranges & $\alpha,\beta\,=\,0,1,2,3$; $i,j\,=\,1,2,3$\\
Flat metric &$\eta_{\alpha \beta}=$ diag$(1,-1,-1,-1)$\\
Coordinates &$x^\alpha=(x^0,x^1,x^2,x^3)=(ct,x^1,x^2,x^3)$\\
Vectors & \b{v} = $(v^1,v^2,v^3)$; $\nabla=(\partial/\partial x^1,\partial/\partial x^2,\partial/\partial x^3) $ \\
Symmetrization & $T_{(\alpha | \beta \dots \gamma |
\delta)}=\frac{1}{2}(T_{\alpha \beta \dots
\gamma\delta}+T_{\delta \beta\dots \gamma\alpha})$ \\
Kronecker &  $\delta^\alpha_\beta$ = 1 if $\alpha = \beta$, $0$ else\\
Connection & $\Gamma^\alpha_{\mu \nu}=\frac{1}{2}g^{\alpha\sigma}\left( g_{\mu\sigma,\nu}+g_{\nu\sigma,\mu}-g_{\mu\nu,\sigma}\right)$\\
Riemann tensor & $R^{\alpha }{}_{\beta \mu \nu } =\Gamma _{\beta
\nu ,\mu }^{\alpha }-\Gamma _{\beta \mu ,\nu }^{\alpha}+\Gamma
_{\beta \nu }^{\sigma }\Gamma _{\sigma \mu}^{\alpha
}-\Gamma_{\beta \mu }^{\sigma }
\Gamma_{\sigma \nu }^{\alpha }$ \\
Ricci tensor & $R_{\mu \nu } =R^{\sigma }{}_{\mu \sigma \nu  }$\\
\hline\hline\hline
\end{tabular}

\caption{\label{tab:definitions}List of conventions and
definitions.}

\end{table}

%\newpage

\begin{table}

\begin{tabular}{cl}
\hline \hline
Dimensions of Physical  Quantities\\
\hline
& \\
1 & $g_{\alpha \beta },$ $\delta _{\alpha \beta }$, $\mathcal{G}$, $a_1$, $K_{I,1}$ \\
& \\
m & $x^{\alpha }$, ${\Gamma^\alpha_{\mu\nu}}^{-1}$, ${R_{\alpha
\beta\mu\nu }}^{-\frac{1}{2}}$, ${R_{\alpha \beta
}}^{-\frac{1}{2}}$, $R^{-\frac{1}{2}}$, $Y_I$, $Y^{-1}_{II}$,
$\lambda^{-1}_1$, $\lambda^{-1}_2$, \\ & $\delta^{-\frac{1}{3}}$,
$\tilde{\mathcal{G}}^{\frac{1}{3}}$, $\xi$, ${a_2}^{\frac{1}{2}}$,
${b_1}^{\frac{1}{2}}$, $K_{I,2}$, ${K_{I,3}}^{-1}$,
${K_{I,4}}^{-1}$, ${K_{I,5}}^{-1}$, \\ & ${K_{I,6}}^{-1}$, ${K_{I,7}}^{-2}$, $K_{II,3}$, $K_{II,4}$, $K_{II,5}$, $K_{II,6}$, $r$ \\
& \\
s & $\sigma^{-\frac{1}{2}}_I$, $\sigma^{-\frac{1}{2}}_{II}$, $\tau^{-\frac{1}{2}}_{II}$ \\
& \\
$\rm{Kg}$ & $M$ \\
& \\
$\frac{\rm{m}}{\rm{s}}$ & $c$, $U^{\frac{1}{2}}$,
$V^{\frac{1}{2}}$, $h_{0 a }^{\frac{1}{3}}$, $v^{a}$, $\Pi
^{\frac{1}{2}}$, $A^{\frac{1}{2}}$, $\Phi^{\frac{1}{2}}$,
$\tau_I^{\frac{1}{2}}$ \\
& \\
$\frac{\rm{m}^{3}}{\rm{s}^{2}}$ & $U_0$, $U_1$, $U_2$ \\
& \\
$\frac{\rm{m}^{4}}{\rm{s}^{2}}$ & $B$ \\
& \\
$\frac{\rm{kg}}{\rm{m}^{3}}$ & $\rho $ \\
& \\
$\frac{\rm{kg}}{\rm{s}^{2}\rm{m}}$ & $T_{\alpha \beta }$, $p$\\
& \\
$\frac{\rm{m}^3}{\rm{s}^{2}\rm{Kg}}$ & $G$\\
\hline\hline
\end{tabular}
\caption{\label{tab:dimensions}Dimensions of the quantities
considered in the paper.}
\end{table}

\end{document}